\documentclass[useAMS,usenatbib,usegraphicx]{mn2e}
\def\msol{\hbox{\kern 0.20em $M_\odot$}}

\newcommand{\lsol}{\hbox{\kern 0.20em $L_\odot$}}
\newcommand{\g}{\hbox{\kern 0.20em g}}
\newcommand{\gmu}{\hbox{\kern 0.20em g$^{-1}$}}
\newcommand{\kg}{\hbox{\kern 0.20em kg}}
\newcommand{\pc}{\hbox{\kern 0.20em pc}}
\newcommand{\mum}{\hbox{\kern 0.20em $\mu$m}}
\newcommand{\mumd}{\hbox{\kern 0.20em $\mu$m$^{-2}$}}
\newcommand{\cm}{\hbox{\kern 0.20em cm}}
\newcommand{\m}{\hbox{\kern 0.20em m}}
\newcommand{\km}{\hbox{\kern 0.20em km}}
\newcommand{\nm}{\hbox{\kern 0.20em nm}}
\newcommand{\s}{\hbox{\kern 0.20em s}}
\newcommand{\h}{\hbox{\kern 0.20em h}}
\newcommand{\smu}{\hbox{\kern 0.20em s$^{-1}$}}
\newcommand{\srmu}{\hbox{\kern 0.20em sr$^{-1}$}}
\newcommand{\smd}{\hbox{\kern 0.20em s$^{-2}$}}
\newcommand{\an}{\hbox{\kern 0.20em an}}
\newcommand{\anmu}{\hbox{\kern 0.20em an$^{-1}$}}
\newcommand{\yr}{\hbox{\kern 0.20em yr}}
\newcommand{\yrmu}{\hbox{\kern 0.20em yr$^{-1}$}}
\newcommand{\Myr}{\hbox{\kern 0.20em Myr}}
\newcommand{\Mymu}{\hbox{\kern 0.20em Myr$^{-1}$}}
\newcommand{\K}{\hbox{\kern 0.20em K}}
\newcommand{\pcmu}{\hbox{\kern 0.20em pc$^{-1}$}}
\newcommand{\pcmd}{\hbox{\kern 0.20em pc$^{-2}$}}
\newcommand{\pcmt}{\hbox{\kern 0.20em pc$^{-3}$}}
\newcommand{\kms}{\hbox{\kern 0.20em km\kern 0.20em s$^{-1}$}}
\newcommand{\kmpd}{\hbox{\kern 0.20em km$^{2}$}}
\newcommand{\kpc}{\hbox{\kern 0.20em kpc}}
\newcommand{\cms}{\hbox{\kern 0.20em cm\kern 0.20em s$^{-1}$}}
\newcommand{\erg}{\hbox{\kern 0.20em erg}}
\newcommand{\ergs}{\hbox{\kern 0.20em erg}}
\newcommand{\cmpd}{\hbox{\kern 0.20em cm$^2$}}
\newcommand{\cmmd}{\hbox{\kern 0.20em cm$^{-2}$}}
\newcommand{\cmms}{\hbox{\kern 0.20em cm$^{-6}$}}
\newcommand{\cmpt}{\hbox{\kern 0.20em cm$^3$}}
\newcommand{\cmmt}{\hbox{\kern 0.20em cm$^{-3}$}}
\newcommand{\mpd}{\hbox{\kern 0.20em m$^2$}}
\newcommand{\mmd}{\hbox{\kern 0.20em m$^{-2}$}}
\newcommand{\mpt}{\hbox{\kern 0.20em m$^3$}}
\newcommand{\mmt}{\hbox{\kern 0.20em m$^{-3}$}}
\newcommand{\mujy}{\hbox{\kern 0.20em $\mu$Jy}}
\newcommand{\mjy}{\hbox{\kern 0.20em mJy}}
\newcommand{\Mj}{\hbox{\kern 0.20em MJy}}
\newcommand{\jy}{\hbox{\kern 0.20em Jy}}
\newcommand{\ghz}{\hbox{\kern 0.20em GHz}}
\newcommand{\G}{\hbox{\kern 0.20em G}}
\newcommand{\muG}{\hbox{\kern 0.20em $\mu$G}}

\usepackage{natbib,graphics,graphicx,subfigure}

\title[Diagnosing shock temperature with NH$_3$ and H$_2$O profiles]{Diagnosing shock temperature with NH$_3$ and H$_2$O profiles}

\author[G\'omez-Ruiz et al.]{A.I. G\'omez-Ruiz$^{1,2}$, C. Codella$^{2}$, S. Viti$^{3}$, I. Jim\'enez-Serra$^{3}$, G. Navarra$^{4}$, \newauthor
R. Bachiller$^{5}$, P. Caselli$^{6}$, A. Fuente$^{7}$, A. Gusdorf$^{8}$, B. Lefloch$^{9}$, A. Lorenzani$^{2}$, B. Nisini$^{10}$  
\\
$^{1}$ CONACYT-Instituto Nacional de Astrof\'isica, \'Optica y Electr\'onica, Luis E. Erro 1, 72840 Tonantzintla, Puebla, M\'exico\\
$^{2}$ INAF, Osservatorio Astrofisico di Arcetri, Largo E. Fermi 5, 50125 Firenze, Italy\\
$^{3}$ Department of Physics and Astronomy, University College London, London, UK\\
$^{4}$ Dipartimento di Fisica, Universit\`a degli Studi di Palermo, Viale delle Scienze, 90128 Palermo, Italy\\
$^{5}$ Observatorio Astron\'omico Nacional (OAN, IGN), Alfonso XII, 3, E-28014 Madrid, Spain\\
$^{6}$ Max-Planck-Institüt f\"ur extraterrestrische Physik, Giessenbachstrasse 1, 85748 Garching, Germany\\ 
$^{7}$ Observatorio Astron\'omico Nacional (OAN, IGN), Apdo 112, E-28803 Alcal\'a de Henares, Spain\\
$^{8}$ LERMA, UMR 8112 du CNRS, Observatoire de Paris, \'Ecole  Normale Sup\'erieure, 61 Av. de l'Observatoire, 75014, Paris, France\\
$^{9}$ Univ. Grenoble Alpes, CNRS, Institut de Plan\'etologie et d'Astrophysique de Grenoble (IPAG), 38401 Grenoble, France\\ 
$^{10}$ INAF, Osservatorio Astronomico di Roma, via di Frascati 33, 00040, Monte Porzio Catone, Italy}

\begin{document}

\date{Accepted date. Received date; in original form date}

\pagerange{\pageref{firstpage}--\pageref{lastpage}} \pubyear{2016}

\maketitle

\label{firstpage}

\begin{abstract}

In a previous study of the L1157 B1 shocked cavity, a comparison between NH$_3$(1$_0$-$0_0$) and H$_2$O(1$_{\rm 10}$--1$_{\rm 01}$) transitions showed a striking difference in the profiles, with H$_2$O emitting at definitely higher velocities. This behaviour was explained as a result of the high-temperature gas-phase chemistry occurring in the postshock gas in the B1 cavity of this outflow. If the differences in behaviour between ammonia and water are indeed a consequence of the high gas temperatures reached during the passage of a shock, then one should find such differences to be ubiquitous among chemically rich outflows.  
In order to determine whether the difference in profiles observed between NH$_3$ and H$_2$O is unique to L1157 or a common characteristic of chemically rich outflows, we have performed Herschel-HIFI observations of the NH$_3$(1$_0$-0$_0$) line at 572.5 GHz in a sample of 8 bright low-mass outflow spots already observed in the H$_2$O(1$_{\rm 10}$--1$_{\rm 01}$) line within the WISH KP. We detected the ammonia emission at high-velocities at most of the outflows positions. In all cases, the water emission reaches higher velocities than NH$_3$, proving that this behaviour is not exclusive of the L1157-B1 position. Comparisons with a gas-grain chemical and shock model confirms, for this larger sample, that the behaviour of ammonia is determined principally by the temperature of the gas. 

\end{abstract}


   \maketitle

\begin{table*}
\centering
\caption{Source list and observed positions.}
\label{soulist}
\centering
\begin{tabular}{lrrcccc}
\hline
Source & $\alpha_{\rm J2000}$ & $\delta_{\rm J2000}$ & $V_{\rm LSR}$ & $L_{\rm bol}$ & $d$ & Offsets
Blue/Red \\
 & ($^{\rm h}$ $^{\rm m}$ $^{\rm s}$) & ($^{\rm o}$ $\arcmin$ $\arcsec$) & (km s$^{-1}$) & ($L_{\odot}$)
 & (pc) & (arcsec) \\
\hline
L1448          & 03 25 38.9 & +30 44 05 & +4.7 & 6  & 235 & B2(-13,+29), R4(+26,-125) \\ 
NGC1333-IRAS2A & 03 28 55.4 & +31 14 35 & +6.0 & 25 & 235 & B(-100,+25), R(+70,-15) \\ 
NGC1333-IRAS4A & 03 29 10.4 & +31 13 31 & +6.5 & 8  & 235 & B(-6,-19), R(+14,+25) \\ 
L1157          & 20 39 06.2 & +68 02 16 & +2.6 & 4  & 250 & B2(+35,-95), R(-30,+125) \\ 
\hline
\end{tabular}
\begin{center}
Note-- For references to the coordinates and protostellar properties see Tafalla et al. 2013; for the distance estimation see \citet{Looney07} and 
\citet{Hirota08}. \\ 
\end{center}
\end{table*}

\section{Introduction}

A newborn protostar generates a fast and well collimated jet, possibly surrounded by a wider angle wind. In turn, the ejected material drives (bow-)shocks travelling through the surrounding high-density medium and traced by H$_2$ ro-vibrational lines at excitation temperatures of around 2000 K. Consequently, slower and cold (10--20 K) molecular outflows are formed by swept-up material, usually traced by CO. Shocks heat the gas and trigger several processes such as endothermic chemical reactions and ice grain mantle sublimation or sputtering. Several molecules, such as H$_2$O, NH$_3$, CH$_3$OH, H$_2$CO, undergo spectacular enhancements by orders of magnitude in their abundances \citep{vandis98}, as observed at mm-wavelengths in a number of outflows \citep{Garay98,Bachiller97,Jorgensen07}. The link between the gas components at $\sim$ 10 K and the hot 2000 K shocked component is crucial to understand how the protostellar wind transfers momentum and energy back to the ambient medium. In this context, studies of the chemical composition of typical molecules in bow-shocks are essential because they represent a very powerful diagnostic tool for probing their physical conditions. Such studies are also paramount to get a comprehensive understanding of chemistry throughout the various phases of the interstellar medium.

    \begin{figure}
   \centering
      \includegraphics[bb=82 27 537 450,angle=-90,width=9.0cm]{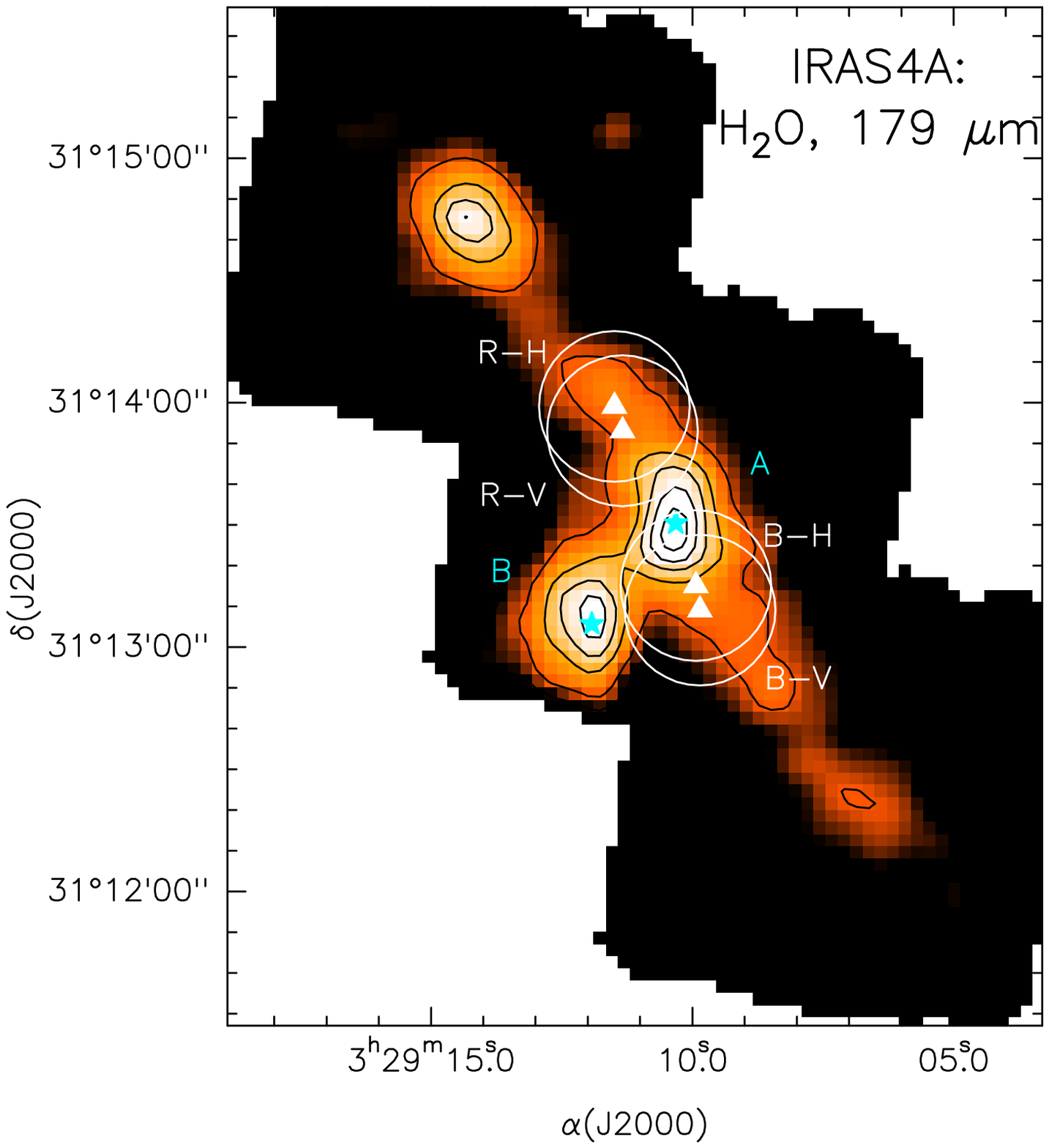}
      \caption{NGC 1333 IRAS4A outflow and corresponding observed positions. The background image represent the H$_2$O emission at 179 $\mu$m from the WISH program \citep{Santangelo14}. The circles show the vertical (V) and horizontal (H) polarization HPBW at the observed positions, whose centers are indicated by the triangles. The stars mark the position of the continuum sources \citep[A and B:][]{Looney00}.  
              }
         \label{map-iras4}
   \end{figure}

    \begin{figure}
   \centering
      \includegraphics[bb=72 65 532 428,angle=-90,width=9.0cm]{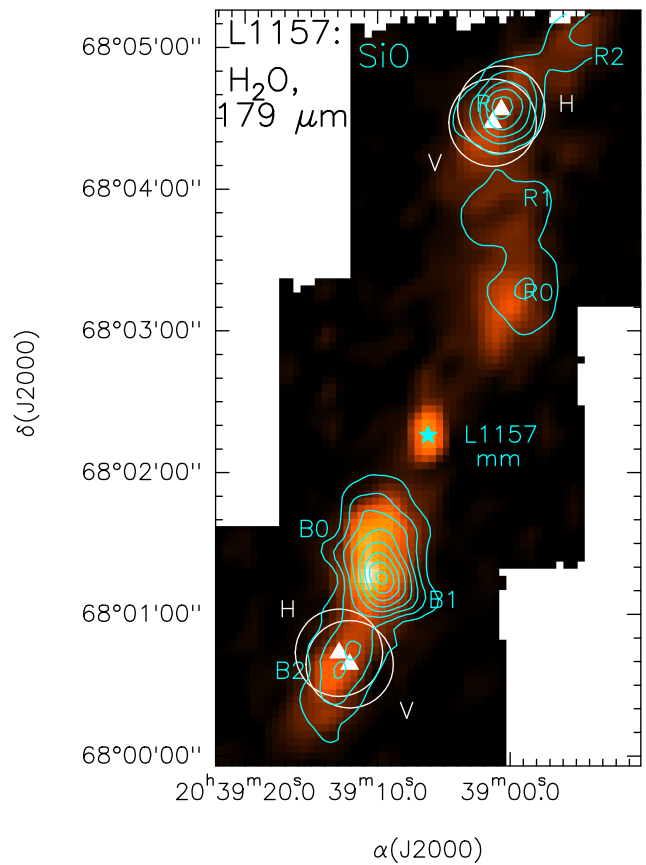}
      \caption{L1157 outflow and corresponding observed positions. The background image represent the H$_2$O emission at 179 $\mu$m from the WISH program \citep{Nisini10b}. The blue contours show the SiO emission from \citet{Bachiller2001}. The circles show the vertical (V) and horizontal (H) polarization HPBW at the observed positions, whose centers are indicated by the triangles. The star marks the position of the central source (L1157-mm). 
              }
         \label{map-l1157}
   \end{figure}

As part of the Herschel Key Program CHESS \citep[Chemical Herschel Surveys of Star forming regions:][]{Cecc10}, the bow-shock L1157-B1 has been investigated with a spectral survey using the HIFI instrument. From the comparison between NH$_3$(1$_0$--0$_0$) and H$_2$O(1$_{\rm 10}$--1$_{\rm 01}$) profiles, a straightforward estimate of the relative abundance ratios of the gas at different velocities was obtained \citep{Codella10}. As a notable example, the NH$_3$/H$_2$O intensity ratio decreases by a factor of $\sim$ 5 moving towards higher velocities suggesting, in case of optically thin emission along the wings, a similar decrease in the abundance ratios. Other tracers of shocked material such as CH$_3$OH and H$_2$CO show the same profile as that of NH$_3$. In \citet{Codella10} we propose that the difference between the H$_2$O and other species reflects different formation mechanisms: for example, while the bulk of NH$_3$ is released from the grain mantles, H$_2$O is enhanced by the release of the icy mantles {\it as well as} by endothermic reactions occurring in the warm ($\geq 220$ K) shocked gas, which convert all gaseous atomic oxygen into water \citep[e.g.,][and references therein]{Isaskun08}. 
However, a model by \citet{Viti11} made especially for these data set suggests that the differences observed in the profile of the different molecular tracers are due mainly to the temperature of the gas: if the latter undergoes a period at a temperature close to 4000 K, then NH$_3$ is easily destroyed by the reaction with hydrogen which leads to NH$_2$ $+$ H$_2$ (this reaction has a high barrier of $\sim$5000 K), 
while H$_2$O remains high in abundance. Such scenario can be explained by the presence of a C-type shock whose pre-shock density and velocity are such that the maximum temperature of the shock reaches 4000 K along the B1 shock of L1157. These findings called for observations of more molecular shocked regions associated with protostellar outflows to investigate whether the difference in profiles between H$_2$O and other species are unique to L1157 or whether it is an ubiquitous characteristic of chemically rich outflows.


    \begin{figure}
   \centering
      \includegraphics[bb=152 67 511 342,angle=-90,width=9.0cm]{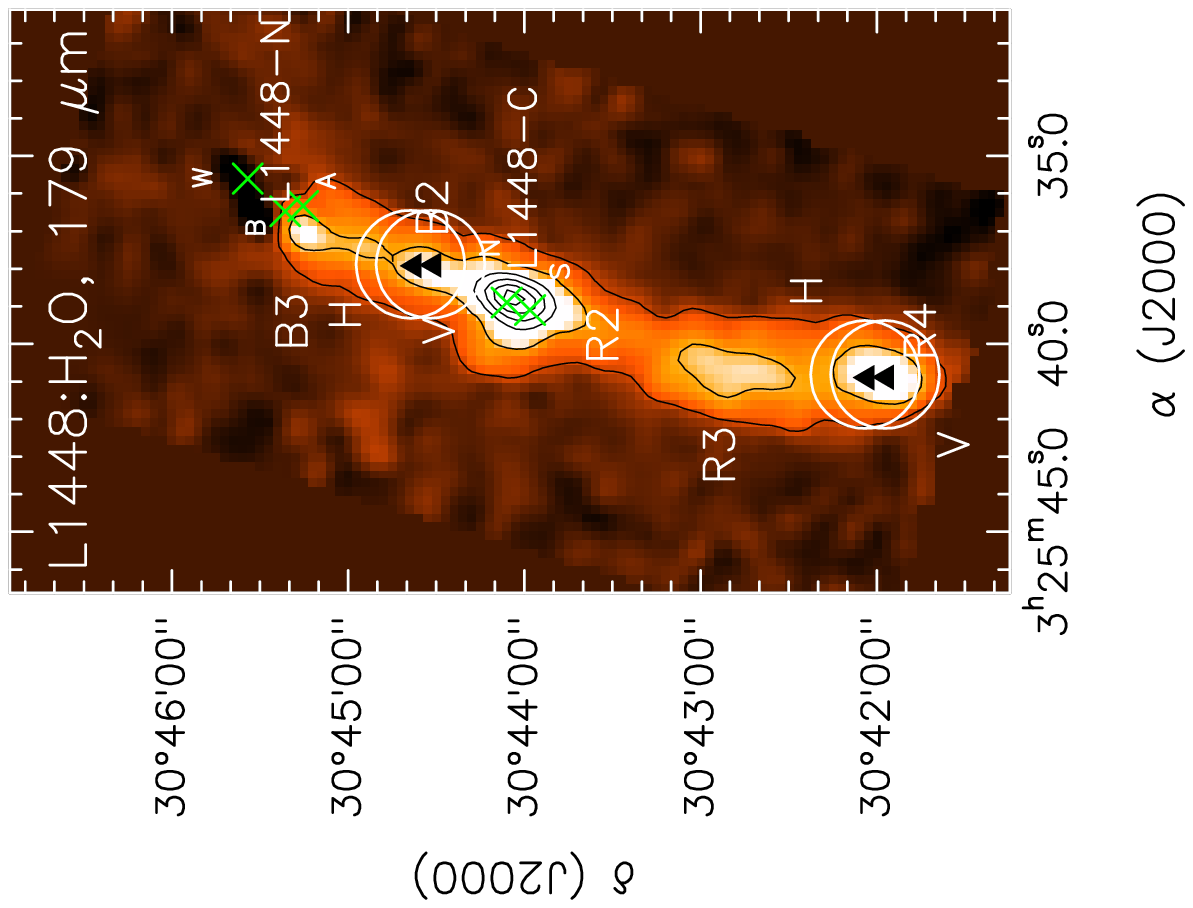}
      \caption{L1448 outflow and corresponding observed positions. The background image represent the H$_2$O emission at 179 $\mu$m from the WISH program \citep{Nisini13}. The circles show the vertical (V) and horizontal (H) polarization HPBW at the observed positions, whose centers are indicated by the triangles. The green crosses indicate the positions of the millimeter continuum sources \citep[N, S, A, B, and W:][]{Kwon06}.  
              }
         \label{map-l1448}
   \end{figure}   

In this article we present observations of the $J_K=$1$_{\rm 0}$--0$_{\rm 0}$ transition of ortho-NH$_3$ at 572.5 GHz in a number of outflow spots already observed in the ortho-H$_2$O(1$_{\rm 10}$--1$_{\rm 01}$) line as part of the Herschel Key Program WISH \citep[Water In Star-forming regions with Herschel:][]{WISH} and reported by \citet{Tafalla13}. In Sect. 2 the target selection and Herschel observations with HIFI are described, in Sect. 3 we report the line profiles obtained, in Sect. 4 and 5 we develop the analysis of the data, and in Sect. 6 we present the summary and conclusions.

    \begin{figure}
   \centering
\includegraphics[angle=90,width=8.5cm]{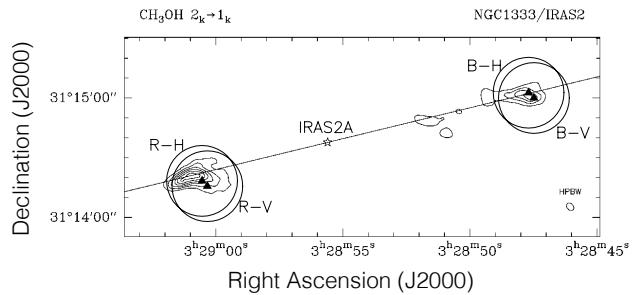}
      \caption{NGC1333 IRAS2A outflow and corresponding observed positions \citep[image adapted from][]{Bachiller98}. The contours show the CH$_3$OH ($2_k-1_k$) PdBI map. The circles show the vertical (V) and horizontal (H) polarization HPBW at the observed positions, whose centers are indicated by the triangles. The star indicates the position of the central source (IRAS2A). 
              }
         \label{map-iras2}
   \end{figure}

\section{Observations}

\subsection{Targeted outflow positions}

The WISH protostellar outflow sample contains 26 outflows driven by Class 0/I low-mass Young Stellar Objects \citep{Tafalla13}. For each source, two ’hot’ spots (blue- and red-shifted) have been observed. The positions were carefully selected by inspecting the maps obtained from ground-based facilities using typical tracers of shocked material (e.g. high-velocity CO, CH$_3$OH, SiO, H$_2$CO). The selected spots are associated with a definite enhancement of the abundance of molecules tracing shocked material or with high-velocity molecular components (up to about 100 km s$^{-1}$ with respect to the systemic velocity). 

From the WISH survey we selected a sample containing 8 'hot' spots in shocks associated with four 'classical' flows driven by low-mass Young Stellar Objects (YSOs) in the earliest evolutionary stages (Class 0), and associated with bright H$_2$O emission: L1448, L1157, NGC1333-IRAS2A, NGC1333-IRAS4A (see Table \ref{soulist}). These YSOs are approximately at the same distance from Earth, with bolometric luminosities between 4-25 $L_{\rm bol}$. 

\subsection{HIFI Observations}

The observations were carried out with the band 1b of the HIFI instrument on-board Herschel, during 2011 October and December, and 2012 February, as part of an OT-1 observing program (OT1\_ccodella\_1). The observations log is shown in Table \ref{obs}. The pointed positions are shown in Table 1 and indicated on Figs. 1--4. The single-pointing observations were made in Position Switching mode and with spatial offsets derived by inspecting large-scale CO maps in order to avoid off-source contamination.

\begin{table}
\centering
\caption{HIFI band 1b observations.}
\label{obs}
\centering
\begin{tabular}{lccc}
\hline
Target & Duration & Date & Obs. Id \\
&(seconds)&&\\
\hline
L1448-R4 & 5703 & 2012-02-24 & 1342239631 \\
L1448-B2 & 5703 & 2012-02-24 & 1342239632 \\
IRAS4A-B & 5703 & 2012-02-25 & 1342239633 \\
IRAS4A-R & 5703 & 2012-02-25 & 1342239634 \\
IRAS2A-B & 5703 & 2012-02-25 & 1342239635 \\
IRAS2A-R & 5703 & 2012-02-25 & 1342239636 \\
L1157-R & 12828 & 2011-12-22 & 1342235065 \\
L1157-B2 & 5703 & 2011-10-08 & 1342230366 \\
\hline
\end{tabular}
\end{table}

The Wide Band Spectrometer (WBS) was used with a frequency resolution of 1.1 MHz. Two High Resolution Spectrometers (HRS) were used in parallel with a frequency resolution of 250 kHz. The observing set-up was prepared in order to observe the o--NH$_3$(1$_0$--0$_0$) line in lower side band with both the WBS as well as with one HRS unit. In addition to our target line, we observed the CS(12--11) line at 587.6 GHz in the upper side band with the WBS and the second HSR unit, and the CO(5--4) line at 576.3 GHz in lower side band with the WBS. Also, the a--CH$_3$OH (12$_{1,11}$--11$_{1,10}$) line at 584.8 GHz and the p--H$_2$CO (8$_{2,6}$--7$_{2,5}$) at 587.4 GHz were observed in the upper side band with the WBS only. Both H and V polarization were observed and then averaged to increase sensitivity. However, we caution that H and V pointings were separated by 6$\farcs$6 
and therefore each polarization may cover a slightly different region. Thus our analysis refers to the whole area covered by the two beams. In appendix A we provide and discuss the H and V spectra. The molecular line parameters as well as Herschel's antenna HPBW, taken according to \citet{Roelf12}, are reported in Table \ref{nspecie}. The Herschel observations were processed with the ESA-supported package HIPE 8.10\footnote{HIPE is a joint development by the Herschel Science Ground Segment Consortium, consisting of ESA, the NASA Herschel Science Center, and the HIFI, PACS and SPIRE consortia.} \citep[Herschel Interactive Processing Environment;][]{Ott10}. FITS files from level 2 were then created and transformed into GILDAS\footnote{http://iram.fr/IRAMFR/PDB/gildas/gildas.html} format for data analysis. 

The spectra in this article are reported in units of main-beam brightness temperature (T$_{\rm MB}=$T$_A^* \times$ $F_{\rm eff}$/$B_{\rm eff}$), for which we have used the $F_{\rm eff}$ of 0.96 and $B_{\rm eff}$ of 0.76 (nominal for band 1b) for all the lines, according to Roelfsema et al. (2012). After smoothing, the spectral resolution in all the cases is 0.5 km s$^{-1}$. 

\begin{table}
\centering
\caption{Transitions and parameters of the lines observed.}
\footnotesize{
\label{nspecie}
\centering
\begin{tabular}{lcccc}
\hline
Transition$^a$ & $\nu_{\rm 0}$ & $E_{\rm u}$ & $A_{\rm ij}$ & HPBW \\
           & (GHz)            & (K) & (10$^{-3}$s$^{-1}$)&($''$)  \\
\hline
o--NH$_3$(1$_0$--0$_0$)            & 572.498 & 28 & 1.61 & 37 \\
CO(5--4)                          & 576.267  & 83 & 0.01 & 37 \\
CH$_3$OH(12$_{1,11}$--11$_{1,10}$)A$^+$  & 584.822  & 197 & 0.89 & 36 \\
p--H$_2$CO(8$_{2,6}$--7$_{2,5}$)      & 587.453   & 173 & 5.66 & 36 \\
CS(12--11)                        & 587.616   & 183  & 4.34 & 36 \\
\hline
o--H$_2$O(1$_{\rm 10}$--1$_{\rm 01}$)$^b$ & 556.936   & 61  & 3.46 & 39 \\
\hline
\end{tabular}}
\begin{center}
$^a$ Transition properties are taken from the Cologne Database for Molecular Spectroscopy: \citet{muller05}. $^b$ Taken from \citet{Tafalla13} \\
\end{center}
\end{table}


\section{Results}

\begin{table*}
\centering
\caption{Line properties.}
 \scriptsize{
\label{linesdata}
\centering
\begin{tabular}{lccccc}
\hline
Transition$^a$ & $T_{\rm MB}^{\rm peak}$ & r.m.s. & $V_{peak}$ & $V_{min},V_{max}$$^b$ & $\int$ $T_{\rm MB}dv$$^c$\\
           & (mK) & (mK) & (km s$^{-1}$) & (km s$^{-1}$) & (mK km s$^{-1}$)\\
\hline
\multicolumn{6}{c}{IRAS4A-B}\\
\hline
o--NH$_3$(1$_0$--0$_0$)    & 280(4)  & 4 & +5.7(0.5) & $-$9.0, $+$12.7 & 624(10) \\
o--H$_2$O(1$_{\rm 10}$--1$_{\rm 01}$)    & 852(2) & 2 & +1.1(0.5) & $-$15.9, $+$25.1 & 14237(9) \\
CO(5--4)           & 5131(2)  & 2 & +6.0(0.5) & $-$18.1, $+$49.7 &  29472(12) \\
CH$_3$OH(12$_{1,11}$--11$_{1,10}$) A$^+$  & 30(3) & 3  & +6.2(0.5) & $+$0.9, $+$9.6 &  130(7) \\
p--H$_2$CO(8$_{2,6}$--7$_{2,5}$)   & 20(3) & 3  & +5.7(0.5) &  $+$3.0, $+$7.8 & 88(7) \\
CS(12--11)           & 46(3) & 3 & +3.3(0.5)  & $-$5.6, $+$13.3 & 532(9) \\
\hline
\multicolumn{6}{c}{IRAS4A-R}\\
\hline
o--NH$_3$(1$_0$--0$_0$)    & 220(4) & 4 & +5.5(0.5) & $-$4.8, $+$24.2 &  721(8) \\
o--H$_2$O(1$_{\rm 10}$--1$_{\rm 01}$)  & 532(2)  & 2 & +10.5(0.5) & $-$1.7, $+$31.9 &  7263(8) \\
CO(5--4)           & 4604(2)  & 2 & +6.0(0.5) & $-$12.3, $+$51.5 &  28245(12) \\
CH$_3$OH(12$_{1,11}$--11$_{1,10}$) A$^+$  & 30(4)  & 4 & +8.0(0.5) & $+$4.9, $+$12.0 &  130(8) \\
p--H$_2$CO(8$_{2,6}$--7$_{2,5}$)   & 30(3)  & 3 & +7.8(0.5) & $+$6.1, $+$10.1 &  77(9) \\
CS(12--11)           & 51(3) & 3 & +8.2(0.5) & $+$5.0, $+$19.4 &  379(10) \\
\hline
\multicolumn{6}{c}{IRAS2A-B}\\
\hline
o--NH$_3$(1$_0$--0$_0$)    & 50(4)  & 4 & +8.5(0.5) & $-$5.0, $+$9.0 &  413(11) \\
o--H$_2$O(1$_{\rm 10}$--1$_{\rm 01}$)  & 565(2)  & 2 & +0.2(0.5) & $-$11.2, $+$7.6 &  6427(6) \\
CO(5--4)            & 1063(3)  & 2 & +5.5(0.5) & $-$16.1, $+$6.2 &  6427(7) \\
CH$_3$OH(12$_{1,11}$--11$_{1,10}$) A$^+$  & -- &  3 & -- & -- & -- \\
p--H$_2$CO(8$_{2,6}$--7$_{2,5}$)   & --   & 3 & -- & -- & -- \\
CS(12--11)            & -- & 2 & -- & --& -- \\
\hline
\multicolumn{6}{c}{IRAS2A-R}\\
\hline
o--NH$_3$(1$_0$--0$_0$)    & 190(4) & 4 &  +6.5(0.5) & $+$4.5, $+$19.2 & 858(14) \\
o--H$_2$O(1$_{\rm 10}$--1$_{\rm 01}$)  & 581(2) & 2 & +11.5(0.5) & $+$3.2, $+$28.5 &  7182(8) \\
CO(5--4)           & 5832(3) & 3 & +6.7(0.5) & $-$11.8, $+$33.3 &  33186(9) \\
CH$_3$OH(12$_{1,11}$--11$_{1,10}$) A$^+$  & 20(3) & 3 & +10.0(0.5) & $+$8.3, $+$10.7 &  64(7) \\
p--H$_2$CO(8$_{2,6}$--7$_{2,5}$)   & --  & 4 & -- & -- &  -- \\
CS(12--11)           & 9(3)  & 3  & +10.0(0.5) & $+$7.2, $+$11.4 &  44(4) \\
\hline
\multicolumn{6}{c}{L1448-B2}\\
\hline
o--NH$_3$(1$_0$--0$_0$)    & 120(4) & 4 & +4.2(0.5) & $+$2.7, $+$6.3 &  295(8) \\
o--H$_2$O(1$_{\rm 10}$--1$_{\rm 01}$)  & 348(2) & 2 & $-$1.1(0.5) & $-$61.7, $+$5.5 &  8911(11) \\
CO(5--4)           & 2711(3) & 3 & +3.5(0.5) & $-$77.3, $+$35.8 &  31603(9) \\
CH$_3$OH(12$_{1,11}$--11$_{1,10}$) A$^+$  & --  & 3  & -- & -- &  -- \\
p--H$_2$CO(8$_{2,6}$--7$_{2,5}$)   & -- & 4  & -- & -- & -- \\
CS(12--11)           & -- & 3 & -- & -- & -- \\
\hline
\multicolumn{6}{c}{L1448-R4}\\
\hline
o--NH$_3$(1$_0$--0$_0$)    & 20(3) & 4 & +10.2(0.5) & $+$6.3, $+$18.8 &  179(14) \\
o--H$_2$O(1$_{\rm 10}$--1$_{\rm 01}$)  & 406(1) & 1 & +19.5(0.5) & $+$1.3, $+$47.4 & 11388(6) \\
CO(5--4)           & 1936(03) & 3 & +6.2(0.5) & $-$1.1, $+$51.5 &  18381(10) \\
CH$_3$OH(12$_{1,11}$--11$_{1,10}$) A$^+$  & --  & 4 & -- & -- &  -- \\
p--H$_2$CO(8$_{2,6}$--7$_{2,5}$)   & --  & 4 & -- & -- & -- \\
CS(12--11)           & -- & 4 & -- & -- &  -- \\
\hline
\multicolumn{6}{c}{L1157-B2}\\
\hline
o--NH$_3$(1$_0$--0$_0$)    & 130(4) & 4 & +1.1(0.5) & $-$4.5, $+$5.5 &  717(9) \\
o--H$_2$O(1$_{\rm 10}$--1$_{\rm 01}$)  & 1150(5) & 5 & +4.5(0.5) & $-$6.6, $+$9.4 &  9274(13) \\
CO(5--4)           & 6394(2) & 2 & +1.9(0.5) & $-$15.2, $+$9.1 &  21733(7) \\
CH$_3$OH(12$_{1,11}$--11$_{1,10}$) A$^+$  & -- & 4 & -- & -- &  -- \\
p--H$_2$CO(8$_{2,6}$--7$_{2,5}$)   & -- & 3 & -- & -- &  -- \\
CS(12--11)           & --  & 4 & -- & -- & -- \\
\hline
\multicolumn{6}{c}{L1157-R}\\
\hline
o--NH$_3$(1$_0$--0$_0$)    & 40(4)& 4 & +7.6(0.5) & $+$2.2, $+$18.7 &  383(13) \\
o--H$_2$O(1$_{\rm 10}$--1$_{\rm 01}$)  & 360(5) & 5 & +19.3(0.5) & $-$1.1, $+$30.5 &  6651(20)\\
CO(5--4)           & 2323(2) & 2 & +4.1(0.5) & $-$2.9, $+$33.5 &  21975(8) \\
CH$_3$OH(12$_{1,11}$--11$_{1,10}$) A$^+$  & -- & 3 & -- & -- &  -- \\
p--H$_2$CO(8$_{2,6}$--7$_{2,5}$)   & -- & 2 & -- & -- &  -- \\
CS(12--11)           & -- & 4 & -- & -- &  -- \\
\hline
\end{tabular}}
\begin{center}
$^a$ Apart from the transitions reported in Table \ref{nspecie}, included here is the o--H$_2$O transition taken from \citet{Tafalla13}. $^b$ Velocity boundaries where the emission is $\geq$ 3$\sigma$. $^c$ The integrated area between $V_{min}$ and $V_{max}$. \\
\end{center}
\end{table*}

Table \ref{linesdata} summarizes the results of the observations, indicating the line intensities ($T_{\rm MB}^{\rm peak}$), velocity of the peak ($V_{\rm peak}$), velocity limits of the emission ($V_{min},V_{max}$), and the total integrated emission ($\int$ $T_{\rm MB}dv$). Table \ref{linesdata} also reports the same parameters for the H$_2$O line, which we measured from the spectra reported in \citet{Tafalla13}. We computed the column densities, from the total integrated emission, assuming LTE and optically thin emission. Table \ref{coldens} shows the results assuming a typical range of temperatures observed toward these kind of objects \citep[20-100 K, e.g.,][]{Lefloch12,Tafalla13}. We find that the ammonia column densities are in the range of $\sim$ 10$^{10}$--5$\times$10$^{11}$ cm$^{-2}$. To probe that these estimations are accurate, in Appendix B we present the results of radiative transfer calculations showing that the ammonia emission is optically thin, $\tau$ $\leq$ 1.3 $\times$ 10$^{-2}$, for these values of column densities and temperatures (see Fig. \ref{tau-nh3}).  For CS, CO, CH$_3$OH, and H$_2$CO, the column densities ranges are $\sim$ 10$^{12}$--5$\times$10$^{14}$ cm$^{-2}$, 10$^{15}$--7$\times$10$^{16}$, 8$\times$10$^{12}$--8$\times$10$^{15}$, and 9$\times$10$^{11}$--9$\times$10$^{13}$ cm$^{-2}$, respectively. Note that in the case of CO, due to the strong absorption, the numbers are only lower limits. 

\subsection{NH$_3$ Profiles}

    \begin{figure*}
   \centering
      \includegraphics[angle=-90,width=15.5cm]{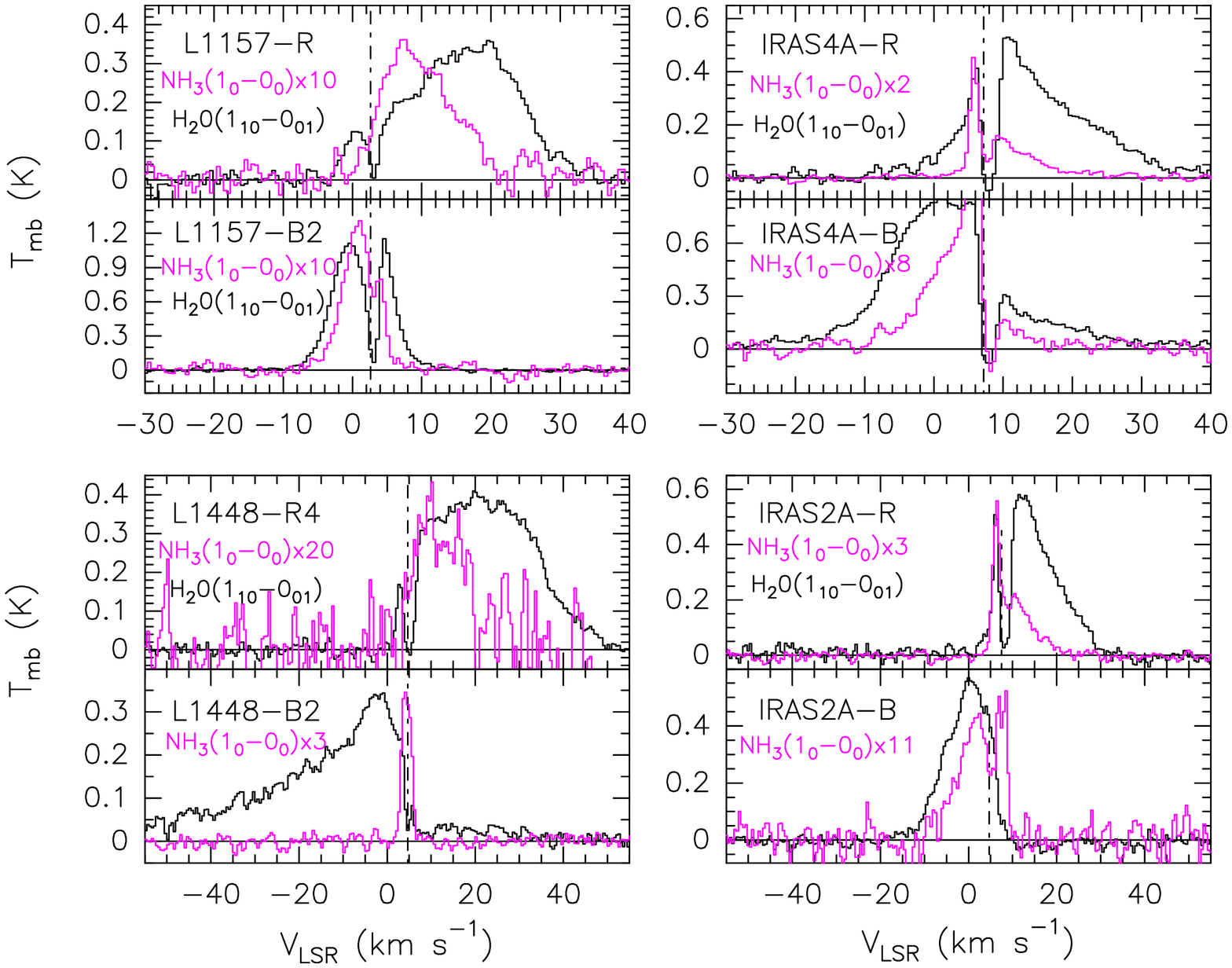}
      \caption{The WBS NH$_3$ spectra (magenta) at the different observed positions. Also displayed are the ortho--H$_2$O spectra (black) from \citet{Tafalla13}. Dashed vertical lines indicates the cloud velocity. 
              }
         \label{nh3-h2o}
   \end{figure*}

The ammonia WBS spectra observed at each of the outflow positions are shown in Fig. \ref{nh3-h2o}, overlaid on the corresponding water spectra. Ammonia emission was detected in all positions, showing extended line wings (up to $\sim$ 15 km s$^{-1}$ with respect to the systemic velocity) in all but the L1448-B2 position, where we found just a narrow profile (a peculiar feature discussed in Sect. 4). 
With the exception of L1448-B2, the WBS spectra show an absorbing dip at systemic velocity plausibly due to the absorption from the extended envelope of the protostar. 

\begin{figure}
   \centering
      \includegraphics[angle=0, bb=1 1 650 360,width=8.5cm]{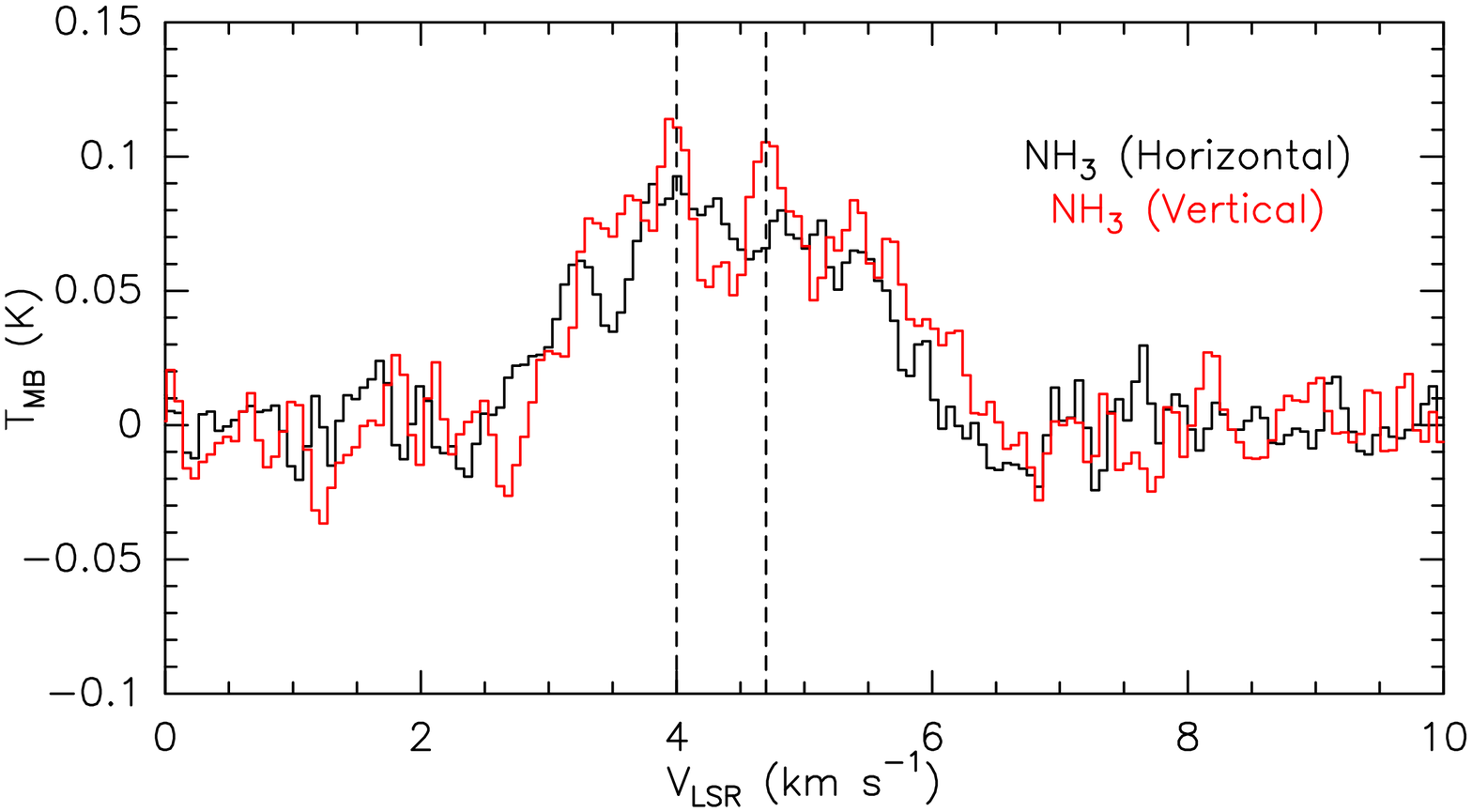}
      \caption{High spectral resolution (HRS) NH$_3$ spectrum towards L1448-B2. Horizontal and Vertical polarization are shown in black and red, respectively. The vertical dashed lines indicate the velocities of the two clouds (4.0 km s$^{-1}$ and 4.7 km s$^{-1}$) found in ammonia centimeter transitions by Bachiller \& Cernicharo (1986).
              }
         \label{L1448B2-hr}
   \end{figure}

The high resolution spectra, from the HRS, do not show additional information on the low-velocity NH$_3$ emission already provided by the WBS spectra, with the exception of L1448-B2, which shows multiple peaks at low velocities (Fig. \ref{L1448B2-hr}). The Horizontal and Vertical polarization spectra are slightly different, as the result of the different area covered by each of them (see Sect. 2.2). The two most prominent spectral features are seen in the Vertical polarization spectrum (which correspond to the beam closer to the central region), showing peaks at 4.0 and 4.7 km s$^{-1}$. These spectral features are discussed in Sect. 4 and 5.  

\subsection{CS, H$_2$CO, and CH$_3$OH profiles}

The spectra of all observed lines are presented in Figs. A1--A4. Only in IRAS4A-B and IRAS4A-R the CS, H$_2$CO, and CH$_3$OH transitions were found. These two positions are also the strongest line emitters among the sources studied here. This is possibly due to the chemical richness of this source as reported by previous investigations \citep[e.g.,][]{Wakelam05,Santangelo14}. However, we point out that later might also be due to sensitivity. We notice that these transitions have larger column densities than NH$_3$. In general, we see that the CS, H$_2$CO, and CH$_3$OH transitions have profiles more similar to NH$_3$ than to H$_2$O, in particular in terms of the maximum velocity. While a more detailed analysis of such species is out of the scope of the present paper, we shall briefly discuss these species in Section 5.

    \begin{figure*}
   \centering
      \includegraphics[angle=-90,width=15.5cm]{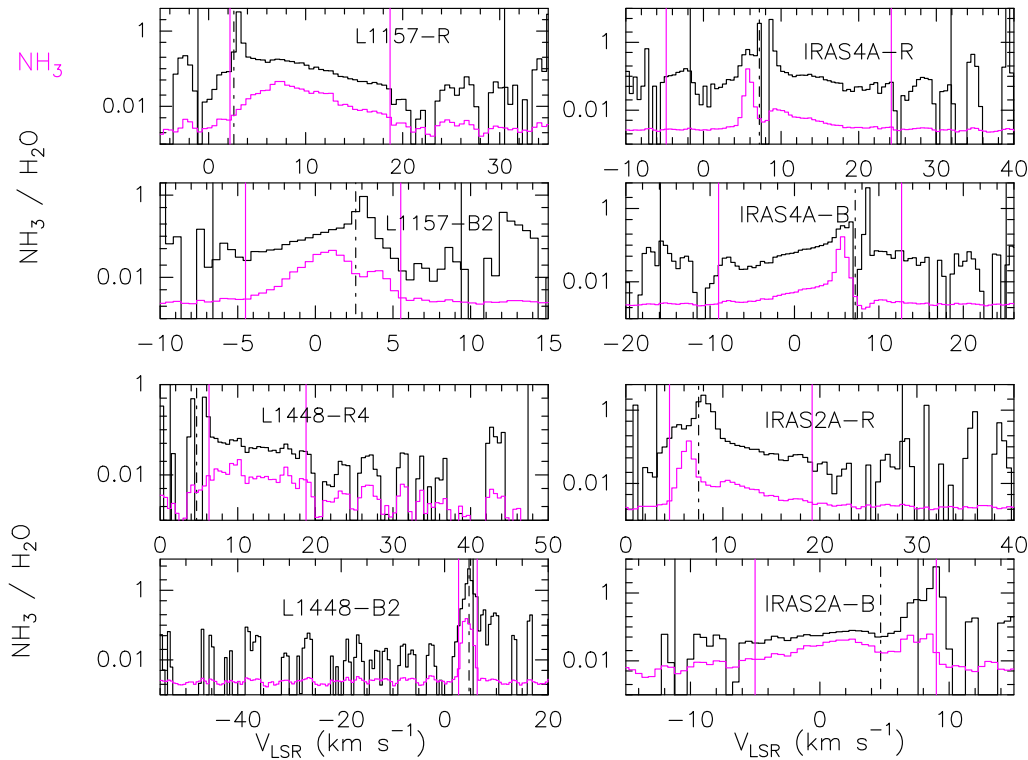}
      \caption{The NH$_3$/H$_2$O intensity ratio at the different observed positions. Dashed vertical lines indicate the cloud velocity. To indicate the region where the ratio is valid, the velocity boundaries (V$_{\rm min}$,V$_{\rm max}$) where the NH$_3$ and H$_2$O emission is $\geq$ 3$\sigma$ are indicated by vertical lines, in magenta and black, respectively. The NH$_3$ spectra are shown in magenta as a reference.
              }
         \label{nh3-h2o-rat}
   \end{figure*}

\section{Ammonia versus water}

In order to properly compare the ammonia and water profiles, a more complete radiative transfer analysis of both species would be necessary to discard excitation and opacity effects. Such analysis is out of the scope of the present work, since we only have one NH$_3$ transition. Despite that, we have shown in Appendix B that our NH$_3$ transition is optically thin at the positions here investigated, and \citet{Tafalla13}, based on the analysis of two H$_2$O transitions, provided evidence that the water emission is also optically thin at the same positions. In addition, radiative transfer models from \citet{Tafalla13} revealed a water gas component with a density in the range of 10$^7$--10$^8$ cm$^{-3}$, and therefore close to critical density of our NH$_3$ transition \citep[$\sim$10$^8$ cm$^{-3}$, e.g.][]{Menten10}. All these information give confidence that the line profile comparison presented in the following is appropriate. We also caution in sect. 5 on the need of radiative transfer calculations in the context of the chemical models.

\begin{figure}
\centering
\includegraphics[angle=0,width=8.0cm]{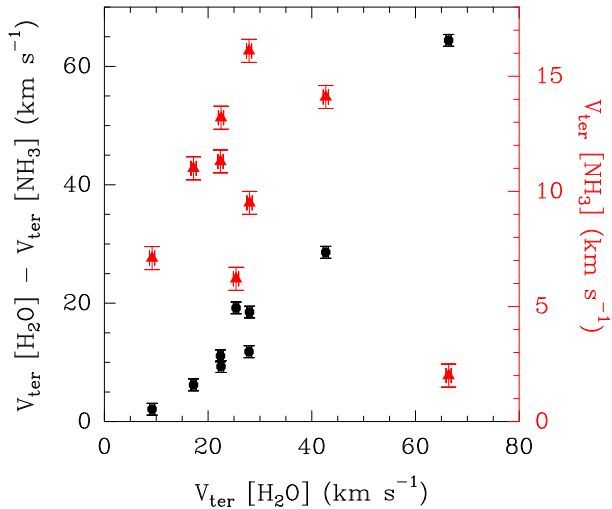}
\caption{Comparison between the maximum (terminal, $V_{\rm ter}$) velocity (absolute values with respect to the systemic velocity, see Table 4) of the H$_2$O emission at 179 $\mu$m and that of NH$_3$ at 572 GHz. Black points show the comparison between $V_{\rm ter}$[H$_2$O]--$V_{\rm ter}$[NH$_3$] versus $V_{\rm ter}$[H$_2$O], while red points compares $V_{\rm ter}$[NH$_3$] with $V_{\rm ter}$[H$_2$O].}  
\label{velocities}
\end{figure}

From the comparison between the NH$_3$ and H$_2$O spectra we see that the profiles of the two species are more often different in shape and in maximum velocity. In all positions the maximum terminal velocity ($V_{\rm ter}$) reached by NH$_3$ is lower than in H$_2$O (always by more than a factor of two; see Fig. \ref{velocities}), which then confirm this tendency first found in L1157-B1 by \citet{Codella10}. The more prominent cases are L1157-R and L1448-B2, which show either a considerably different $V_{\rm peak}$ or very different linewidths. 

The ammonia-to-water line ratio, as function of velocity, is shown in Fig. \ref{nh3-h2o-rat}. The maximum line ratio is found around the cloud velocity in all but IRAS2A-B position. In the L1157-B2 and L1157-R positions the behaviour of the line ratio as a function of velocity is similar to what was found by \citet{Codella10} at the L1157-B1 position, i.e. a maximum close to the cloud velocity and slow decrease of the ratio toward high velocities. 

The most peculiar case is the L1448-B2 position, in which the line ratio profile is just a sharp peak at the systemic velocity. This is the consequence of the very different NH$_3$ line profile with respect to H$_2$O, as pointed out previously. Here the main difference is that while the H$_2$O reaches V$_{\rm LSR}\sim -$ 61 km s$^{-1}$, the NH$_3$ reaches only V$_{\rm LSR}\sim +$3 km s$^{-1}$ (with linewidths of $\sim$ 67 and 4 km s$^{-1}$, respectively). As pointed out in the previous section, the high spectral resolution data (Fig. \ref{L1448B2-hr}) reveal structure of this narrow profile: at least two peaks, at 4.0 and 4.7 km s$^{-1}$.

Similar narrow profiles were also found by \citet{Isaskun05} in CH$_3$OH millimeter transitions at other nearby positions in the blue lobe of L1448, which they interpret as the magnetic precursor \citep[see also][]{Isaskun04,Isaskun09}. However, the spectral feature of such precursor is very narrow ($\sim$ 0.6 km s$^{-1}$) and red-shifted by about 0.5 km s$^{-1}$ with respect to the 4.7 km s$^{-1}$ cloud, i.e. at V$_{\rm LSR}\sim +$5.2 km s$^{-1}$. On the other hand, our NH$_3$ spectra is not that narrow (in fact, a few km s$^{-1}$) and does not show a peak at that V$_{\rm LSR}$. We notice that our peaks are coincident with the peak velocities of the clouds previously reported in centimeter ammonia transitions by \cite{Bachiller86}. Taking into account the discrepancy between the H and V polarization (the two peaks only clearly noticed in the V polarization) and the noise, it does not seem reliable to identify the narrow feature as the magnetic precursor. Although the noisy high resolution spectra might not be sufficient evidence, our ammonia observations of the L1448-B2 position suggest that the emission may be tracing these two clouds enclosed within the HIFI beam. An additional argument for the absence of ammonia emission from the high-velocity shock is that in previous NH$_3$(1,1) and (2,1) interferometric maps by \citet{Curiel99}, no emission from the B2 position was found, with the maps showing emission only towards the central part of L1448-C and L1448-N objects. This peculiar narrow line profile is further discussed, in its comparison with the water profile, within the context of the chemical models in Sect. 5. 



\section{Chemical models}

The NH$_3$-H$_2$O profile differences first discovered in L1157-B1 (e.g. Codella et al. 2010) are clearly present in other outflows. In Viti et al. (2011)
a gas-grain chemical and shock model was used to investigate the differences in profile between the water and the ammonia transitions; it was found that these differences are purely chemical and can be explained by the presence of a C-type shock whose maximum temperature must be close to $\sim$4000 K along the B1 clump. More specifically, assuming, as Codella et al. (2010) imply, that the NH$_3$/H$_2$O intensity ratio decreases at high velocities because a similar decrease in the abundance ratios occurs, then models where the maximum temperature of the shock is close to $\sim$4000K lead to water remaining abundant at all velocities (i.e. throughout the C-shock) while NH$_3$ decreases at high velocities in the warm postshock gas. This is due to i) the efficient destruction of ammonia in the postshock gas via the endothermic reaction H $+$ NH$_3$ $\rightarrow$ NH$_2$ $+$ H$_2$ once its activation barrier ($\sim$5000 K) is overcome at temperatures higher than $\sim$4000 K (see Viti et al. 2011); and ii) to the absence of formation routes of NH$_3$ at high temperature. In Viti et al. (2011) the models that best fit the NH$_3$/H$_2$O also fit the emission from other shock tracers such as CH$_3$OH and H$_2$CO, which fairly follow the behaviour of ammonia (see also Section 3.2 for the case of these molecules in this work). Due to the few detections of the later molecules and the low signal-to-noise in their wings, we do not further discuss them here in the context of the models.  

The maximum temperature of the shock in turn constrains the pre-shock density of the clump and velocity of the shock. If the  temperature of the gas is indeed the key physical parameter determining the amount of ammonia as a function of velocity, then, outflows spots differing in pre-shock density and shock velocities will have different profile behaviours for NH$_3$ and hence different water-to-ammonia ratios as a function of velocities. 

To illustrate this, in Figure 8 we report the difference between the terminal velocities of H$_2$O and NH$_3$ as a function of the terminal velocity of H$_2$O. Since the abundance of H$_2$O remains high across the full shock width (Viti et al. 2011), we use the terminal velocity of the H$_2$O line profile as a proxy of the velocity of the C-shock. We note that this is strictly true for shock velocities between 10 and 45 km$\,$s$^{-1}$, i.e. for the majority of the shock spots discussed in this paper (see Section 5.2 below). Therefore, we assume that the terminal velocity of H$_2$O lies closely to the actual velocity of the shock. As a result, from Figure 8 we find that the terminal velocities of water and ammonia become more unlike for increasing C-shock velocities, suggesting a clear chemical effect for ammonia with the strength of the shock. In this section we test this hypothesis by using the \citet{Viti11} model and by determining a set of physical and shock parameters that can fit each of the spectra.

\subsection{The shock and chemical model}

The code used in \citet{Viti11} is the UCL$\_$CHEM \citep{Viti04a} augmented with a shock module (from Jimenez-Serra et al. 2008). The code runs essentially in two phases: Phase I follows the evolution of a core, with initial density of 100 cm$^{-3}$, gravitationally collapsing; gas-phase chemistry, sticking on to dust particles with subsequent surface processing occur. The sticking coefficients for all species are all assumed to be 1, equivalent to a 100\% sticking efficiency \citep[see][Equation 2]{Rawlings92}. However for consistency with the Viti et al. (2011) study, we have varied these coefficients in order to simulate different degree of freeze out at the end of Phase I. The final density is a free parameter (see below). Note that the collapse in Phase I is not meant to represent the formation of a protostar, but it is simply a way to compute the chemistry of high density gas in a self-consistent way starting from a diffuse atomic gas, i.e. without assuming the chemical composition at its final density. Phase II computes the time-dependent chemical evolution of the gas and dust once the core 
has formed and stellar activity and shocks are present. The model self-consistently accounts for both thermal desorption, due to the dust being heated by the presence of the outflow, as well as sputtering of the icy mantles: the latter are sputtered once the dynamical age across the C-shock has reached the "saturation timescales" ($t_{sat}$; see Table 5), as in Jimenez-Serra et al. (2008). 
In all the cases considered, sputtering occurs at earlier times than thermal desorption. The model is the same as employed in Viti et al. (2011) to model the same shock spot so we refer the reader to that paper for further details. 

For this work, we ran a grid of models varying in (i) pre-shock density (n$_H$); (ii) shock velocity (v$_{\rm shock}$); (iii) efficiency of freeze out of gas phase species during the cold phase (Phase I). The maximum temperature of the shock, which varies depending on the pre-shock density and shock velocity, is extracted from Figures 8b and 9b in \citet{Draine83}. The saturation times are taken from Jimenez-Serra et al. (2008). \footnote{The saturation times are related to the time-scales within the shock at which most of the molecular material in the ices have been injected into the gas phase by the sputtering of dust grains (see Jimenez-Serra et al. 2008 for the actual definition and determination of this parameter
).} Table 5 lists the models ran. Columns 7 lists, for each model, the length of the dissipation region which is the shock length scale and depends on the shock velocity as well as on the pre-shock density.

\begin{table*}
\centering
\caption{Model parameters: Model number, pre-shock density, shock velocity, saturation
time, maximum temperature of the neutral gas (these four parameters are interconnected - see Jimenez-Serra et al. 2008), degree of depletion 
(note that we use the fraction of CO in the icy mantles at the end of Phase I to estimate this fraction), 
and dissipation length. The last column lists the group we categorized the model in (see Sect. 5.1). a(b) stands for a $\times$ 10$^b$.}
\label{models}
\centering
\begin{tabular}{lccccccc}
\hline
n$^o$ & n$_H$ (cm$^{-3}$) & V$_s$ (km s$^{-1}$) & t$_{sat}$ (yrs) & T$_{max}$ (K) & Freeze-out (\%) & L$_{diss}$ (cm) & Group \\ 
\hline
1&10$^5$& 40& 4.6& 4000& 1 & 1.5(16) &2 \\
2&10$^5$& 40& 4.6& 4000& 15 & 1.5(16) &2 \\
3&10$^5$& 40& 4.6& 4000& 30 & 1.5(16) &2 \\
4&10$^5$& 35& 4.6& 3200& 30 & 1.3(16) &1 \\
5&10$^5$& 35& 4.6& 3200& 1 & 1.3(16)  &1\\
6&10$^3$& 40& 455& 2200& 1 & 1.5(18) &1 \\
7&10$^4$& 40& 45.5& 2200& 1 & 1.5(17) &1 \\
8&10$^4$& 60& 38& 4000& 3  & 2.2(17) &2 \\
9&10$^3$& 60& 380& 4000& 3 & 2.2(18) &2 \\
10&10$^3$& 60& 380& 4000& 1 & 2.2(18) &2 \\
11&10$^5$& 40& 4.6& 4000& 60 & 1.5(16) &2 \\
12&10$^4$& 40& 45.5& 2200& 6 & 1.5(17) &1 \\
13&10$^5$& 35& 4.6& 3200& 60  & 1.3(16) &1 \\
14&10$^4$& 10& 10.5& 300& 18 & 3.7(16) &1 \\
15&10$^4$& 40& 45.5& 2200& 18 & 1.5(17) &1 \\
16&10$^3$& 10& 954& 300& 1 & 3.7(17) &3 \\
17&10$^4$& 10& 10.5& 300& 3 & 3.7(16) &1 \\
18&10$^5$& 10& 10.5& 300& 60 & 3.7(15) &1 \\
19&10$^5$& 10& 10.5& 300& 30 & 3.7(15) &1 \\
20&10$^5$& 10& 10.5& 300& 15  & 3.7(15) &1 \\
21&10$^5$& 10& 10.5& 300& 1 & 3.7(15) &1 \\
22&10$^3$& 20& 570& 900& 1 & 7.4(17) &3 \\
23&10$^3$& 20& 570& 900& 3 & 7.4(17) &3 \\
24&10$^4$& 20& 5.7& 900& 3 & 7.4(16) &3 \\
25&10$^4$& 20& 5.7& 900& 18 & 7.4(16) &1 \\
26&10$^4$& 20& 57.0& 900& 18 & 7.4(16) &1 \\
27&10$^4$& 20& 57.0& 900& 3 & 7.4(16) &1 \\
28&10$^5$& 20& 5.7& 800& 15 & 7.4(15) &1 \\
29&10$^5$& 20& 5.7& 800& 1  & 7.4(15) &1 \\
30&10$^5$& 20& 5.7& 800& 60 & 7.4(15) &1 \\
31&10$^5$& 20& 5.7& 800& 30 & 7.4(15) &1 \\
32&10$^4$& 10& 95.4& 300& 18 & 3.7(16) &1 \\
33&10$^4$& 10& 95.4& 300& 3  & 3.7(16) &1 \\
34&10$^6$& 40& 0.5& 4000& 100 & 1.5(15) &1 \\
35&10$^6$& 40& 0.5& 4000& 50 & 1.5(15) &2 \\
36&10$^6$& 40& 0.5& 4000& 80 & 1.5(15) &2 \\
37&10$^3$& 30& 440& 1800& 1 & 1.1(18)  &2 \\
38&10$^6$& 30& 0.4& 2000& 80 & 1.1(15) &3 \\
39&10$^3$& 30& 440& 1800& 3 & 1.1(18) &3 \\
40&10$^4$& 30& 44.0& 1800& 18 & 1.1(17) &1 \\
41&10$^4$& 30& 44.0& 1800& 3 & 1.1(17) &1 \\
42&10$^6$& 30& 0.4& 2000& 100 & 1.1(18) &1 \\
43&10$^6$& 30& 0.4& 2000& 50 & 1.1(18) &1 \\
44&10$^5$& 30& 4.4& 2000& 30 & 1.1(16) &1 \\
45&10$^5$& 30& 4.4& 2000& 1 & 1.1(16) &1 \\
46&10$^5$& 30& 4.4& 2000& 6  & 1.1(16) &1 \\
47&10$^5$& 30& 4.4& 2000& 15 & 1.1(16) &1 \\
48&10$^4$& 15& 68 & 600& 18 &  5.5(16) &1 \\
49&10$^5$& 15& 6.8& 550& 60 & 5.5(15) &1 \\
50&10$^6$& 15& 0.7& 550& 80 & 5.5(14) &1 \\
51&10$^4$& 25& 49 & 1200& 18 & 9.2(16) &1 \\
52&10$^5$& 25& 4.9& 1500& 60 & 9.2(15) &1 \\
53&10$^6$& 25& 0.5& 1500& 80 & 9.2(14) &1 \\
54&10$^4$& 45& 33.8& 2800& 18 & 1.7(17) &1 \\
55&10$^5$& 45& 3.4& 5000& 60 & 1.7(16) &3 \\
56&10$^6$& 45& 0.3& 5000& 80 & 1.7(15) & 3\\
57& 5$\times$10$^4$ & 45& 6.7&  6500 & 30& 3.7(16) &3 \\
58&10$^4$& 65&  38& 10000& 30 & 2.2(17) &3 \\
\hline
\end{tabular}
\end{table*}

\subsection{Chemical trends}
The behaviour of both water and ammonia for each model is analysed and we find that broadly speaking we can divide our models in three groups (see last column of Table 5):
\begin{itemize}
\item Group 1: models where ammonia and water behave in a similar way, i.e. 
they are both either abundant, or otherwise, at each position across the dissipation length; Models 4-7, 12-15, 17-21, 25-34,  40-54 belong to this category. 

\item Group 2: models where ammonia decreases 'earlier' in the postshock gas (i.e at lower velocities) than water; Models 1-3, 8-11, 35-36 belong to this group.

\item Group 3: models where the behaviour of NH$_3$ and/or water does not follow a clear trend. Models 16, 22-24, 37-39, 55-58 belong to this category. 
\end{itemize}
In Figure 9 we plot the fractional abundance of water with respect to the total number of hydrogen nuclei (black line) and ammonia (red line) as a function of velocity within the postshock gas for a selected subset of models covering all the behaviours.

\begin{figure}
   \centering
      \includegraphics[angle=0,width=8.5cm]{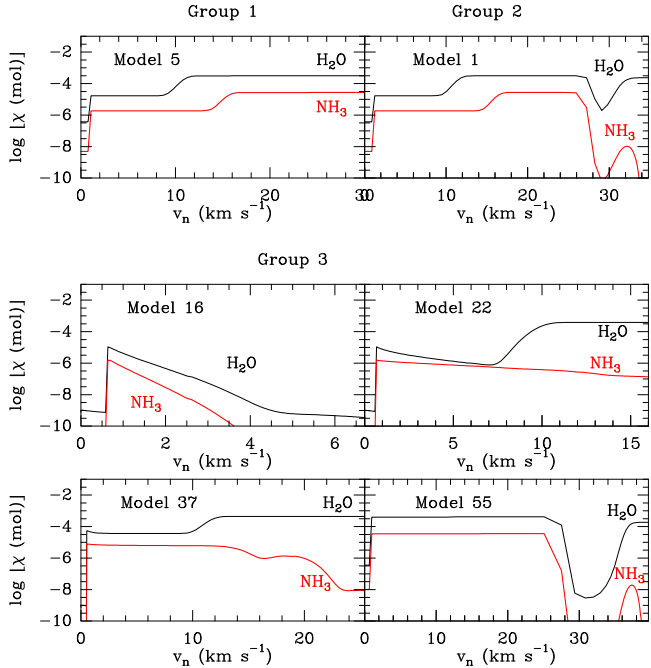}
      \caption{Fractional abundances with respect to the total number of hydrogen nuclei of water and ammonia as a function of velocity within the postshock gas, for representative models within each Group.
              }
         \label{models}
\end{figure}

The behaviour of ammonia and water in 
Groups 1 and 2 were amply discussed in Viti et al. (2011). 
Group 3 includes models where the trend of water and ammonia are not easily categorized; 
for example in Model 16, both species decrease considerably but the water abundance does not drop as much as the ammonia one, and eventually reaches a plateau. Model 22, 23, 24 and 37 have the same behaviour of models belonging to Group 1 up to 7 km s$^{-1}$ but then water increases due to 
the temperature in the shock reaching above $\sim$200 K;
ammonia on the other hand reaches a plateau: both species are in fact quite abundant throughout the dissipation length. 
Models 38 and 39 do show a decrease in ammonia as the velocity increases but without a sharp drop in its abundance, which remains constant at 10$^{-8}$. This decrease is again due to the high gas temperature reached in the shock, which allows the destruction of NH$_3$ to occur (Viti et al. 2011). The decrease of the NH$_3$/H$_2$O ratio for models 22, 23 and 24 is expected to be similar to that for models 38 and 39. However, the reactions responsible for this decrease are different: if the shock velocity is high, ammonia is destroyed (models 38 and 39), while if the shock velocity is low, water is enhanced (models 22, 23 and 24). In Models 55, 56, 57, 58, where the temperature of the gas is $\geq$ 5000 K,  both
water and ammonia decrease in abundance with velocity but then water increases again 
and remains high in abundance, while NH$_3$ only briefly increases in abundance: the higher the 
shock temperature, the larger and longer is the drop in water abundance. 

The subdivision above is purely based on an empirical analysis of our plots. We now attempt at identifying the chemical or physical processes that yield such categorization. In Viti et al (2011) we claimed that the differences between NH$_3$ and H$_2$O are purely chemical and 
are driven by the maximum shock temperature, which needed to be over $\sim$4000~K in order for the NH$_3$ to be destroyed at high velocities. In order to test this claim we now compare the models belonging to Groups 1 and 2 (see Table 5): we note indeed that all models belonging to Group 1 have a maximum shock temperature below 3500~K and, on the contrary, all the models in Group 2 have a temperature of $\sim$4000~K. Of particular interest are the models in Group 3: most have a maximum shock temperature well below 3500~K, indicating that the observed behaviour of our two species is not simply determined 
by whether the maximum temperature of the shock is high enough to destroy ammonia at high velocities but still low enough to not affect the water abundance. It seems that at very low temperatures ($<$ 1000K) the abundance of ammonia never in fact increases in abundance in the first place 
(as its main route of formation, H$_2$ + NH$_2$, is endothermic, with a barrier of $\sim$1400 K - see Viti et al. 2011 for more details on the formation of NH$_3$ under different conditions); more interestingly Models 55-58 have a maximum shock temperature above $\sim$4000~K which means that the clear trend where H$_2$O 
shows bright emission at all the sampled high velocities is only true for a very narrow range of maximum shock temperatures: once the temperature is close to 5000~K water is also destroyed, as indeed we stated in Viti et al. (2011).

In conclusion, while we confirm that it is the maximum shock temperature what determines the behaviour of these species as a function of velocity, different maximum shock temperatures lead to a varied range of behaviours. Independently from the shock model employed, our results indicate that the behaviours of water and ammonia are simply a function of temperature, which is set by the shock velocity and the physical conditions of the pre-shock gas. Although the temperature varies very quickly with the passage of the shock, at the high temperatures reached within it, the chemistry of ammonia is fast enough to leave a signature in the post-shock gas. 

\subsection{Comparison with observations}

Now that we have established the sensitivity of the NH$_3$ and H$_2$O profiles to the shock conditions, we attempt, qualitatively, to associate each observed outflow spot with a Group of models and, if possible, to a range of physical conditions. We note that it is not possible to directly compare the abundances of our models to the observations. Nevertheless if one assumes, as in Codella et al. (2010), that the differences between the H$_2$O and NH$_3$ profiles as a function of velocity should be reflected in the differences in abundances as a function of velocities then one can use the models as plotted in Figure 9 to aid such comparison. 
For the comparison with observations, it is also important to note that the high-temperature region in our shock model does not correspond to the temperature of any gas component that can be observed directly in these outflow spots: what we are likely tracing is the far downstream postshock gas with the 'fingerprint' of the chemistry occurred during the high temperature shock phase(s). We also underline that without a detailed radiative transfer model that takes into consideration the source size, beam dilution and the excitation of the H$_2$O and NH$_3$ lines, it is impossible at this stage to quantitatively match a particular model to an object. In the following, we try to determine the most likely shock parameters that better match the NH$_3$ and H$_2$O line profiles observed in our outflow spot sample. We do this by comparing the observed profile as a function of velocity with the molecular abundance as a function of depth. We recognize that this can only lead to a qualitative match and that the abundances would need to be fed into a radiative transfer model in order to be able to directly fit the observations (as it was done in Viti et al. 2011).

{\it L1157-B2 and IRAS2A-B}: these two objects show a very similar, and relatively narrow, profiles in both molecules; models from Group 1 that seem to best match these objects are Models 45, 50, 53; however they are probably better matched by some Models in Group 3, i.e. Model 16 may be the best match for L1157-B2 while Models 55-57 may be the closest to IRAS2A-B. Model 16 implies that L1157-B2 has a lower pre-shock density than L1157-B1 as well as a much lower shock velocity (10 km s$^{-1}$), which is also consistent with the behaviour of the CO emission (see Figure A.2). IRAS2A-B on the other hand may be an example of a very fast shock, with a high pre-shock density: although the terminal velocity of water is only $\sim$ 20 km/s, the CO spectrum (See Figure A.3) show emission at higher velocities. This object is in fact considered one of the strongest emitters. 
The fact that two very different models are invoked to match two objects with very similar water and ammonia profiles is a consequence of the fact that very different physical and/or chemical conditions can lead to theoretical abundances profiles that can be grouped together. Indeed, the difference in CO profiles between L1157-B2 and IRAS2A-B is an indication that these two objects may in fact be very different.  Nevertheless, it is worth underlining however that since more than one model can match the behavioural trend of the NH$_3$ and H$_2$O, we are {\it not} claiming a unique match between one model and one object.  

{\it L1448-B2, IRAS4A-R, and IRAS2A-R}: these three objects show a narrow ammonia emission, with the water profile being quite extended. We would therefore expect these objects to be best matched by models in Group 2 where ammonia is only abundant for a short period of time (i.e for a small velocity bin). Models that may be good matches are Models 1, 8, 10, where the pre-shock density can range from 10$^3$--10$^6$ cm$^{-3}$  but where the shock velocity is always at least 40 km s$^{-1}$ and the depletion on the grains during the pre-shock phase is low: this implies a lower abundance of water and ammonia at the time of the sputtering of the icy mantles (as both species are enhanced on the grains as a function of freeze out, due to hydrogenation of oxygen and nitrogen respectively). The peculiar narrow profile in L1448-B2 can be then understood in the context of these models: since the NH$_3$ decreases earlier, the narrow line implies that almost all the shocked ammonia is gone and we see mainly the contribution of the cloud cores.  

{\it L1157-R, L1448-R4, IRAS4A-B}: resemble L1157-B1 whereby water is indeed more extended in velocity than ammonia, but the latter does not have a narrow profile. Models from Group 2 are best fits, in particular Models 2, 3, 9, 11, or 35. These models span the same pre-shock densities and maximum velocities as for the L1448-B2, IRAS4A-R, and IRAS2A-R objects but seem to have a higher depletion on grains during Phase I.

The association of particular models to individual objects has been done {\it solely} on the basis of the comparison of the line profiles with the abundances as a function of velocity, as explained at the beginning of this section. The iteration of such parameters to match observations is all we can do without line radiative transfer modelling. It is useful to crudely estimate whether the abundances in our chosen models can at least lead to observable intensities for the ammonia lines. We therefore run some RADEX calculations (van der Tak. et al. 2007) for Models 1, 5 and 16 (as representative of Groups 2, 1, and 3) using representative values of the abundance of NH$_3$ at different velocities. We find that for Models 1 and 5 it is very easy to reach the observed line intensities; for Model 16 we can obtain line intensities of the order of 0.1 K, as long as we use the abundance as averaged only up to 3.5 km s$^{-1}$ and a narrow ($<$ 10 km s$^{-1}$) linewidth.
 
\section{Summary and conclusions}

In the following we summarize the main results and conclusions from our {\it Herschel}/HIFI observations of the ammonia emission from protostellar outflows:

\begin{enumerate}

\item We detected the NH$_3$ emission from all eight outflow positions we have observed. 
In all the cases, the ammonia emission reaches terminal velocities ($V_{\rm ter}$) that are lower than H$_2$O, proving that this behaviour is not exclusive of the L1157-B1 position. 
In addition to ammonia, all the bonus lines (due to CS,
H$_2$CO, and CH$_3$OH) were detected in only IRAS4A-B and IRAS4A-R positions, confirming the chemical richness of these regions.

 \item Comparisons with chemical modelling confirms that the behaviour of ammonia is determined principally by the temperature of the gas. 

\item While a quantitative comparison between models and observations is not feasible without a proper line radiative transfer model, we constrain the pre-shock density and/or shock velocity for each object based on a comparison of abundance trends. We find that, while several models show agreement with the profiles of the different objects, the best matching model for L1157-B2 has a very low pre-shock density (10$^3$ cm$^{-3}$ and velocity (10 km s$^{-1}$), while IRAS2A-B abundances are best reproduced by a gas that has undergone a relatively high velocity shock (45 km s$^{-1}$) with a pre-shock density of $\sim$ 10$^5$ cm$^{-3}$. L1448-B2, IRAS4A-R and IRAS2A-R are matched by models where ammonia is heavily destroyed at high velocities, as explained above due to the short period when the temperature of the gas is high, at 4000 K. We are not able to constrain the pre-shock density for these objects as it can range from as low as 1000 cm$^{-3}$ to as high as 10$^{6}$ {\it as long as} the maximum temperature of the shock is 4000 K, which can be achieved for a shock velocity of $\sim$ 40km s$^{-1}$. The best matching models also indicate a low level of depletion in the cold phase prior to the passage of the shock, hence it is likely that the pre-shock density is in fact towards the lower limit. Finally, L1157-R, L1448-R4 and IRAS4A-B seem to resemble very closely the abundance profile of L1157-B1. They are likely therefore to have a pre-shock density of 10$^5$-10$^6$ cm$^{-3}$ and a shock velocity of the order of 40 km s$^{-1}$, although we can not exclude a faster shock with a lower pre-shock density: in other words, the behaviour of the H$_2$O/NH$_3$ is again determined by the high temperature the gas can attain and the latter can be achieved by more than one combination of shock parameters. In terms of theoretical abundances, a high H$_2$O/NH$_3$ for much of the dissipation length is only reached within a small range of maximum shock temperatures;  in terms of profiles, on the other hand, half of our sample show a high H$_2$O/NH$_3$ ratio: not all of them however require a maximum shock temperature to be close to 4000 K as group 3 models indicate. It is also important to point out that within the observed beam it is unlikely that we are seeing one episodic shock or a group of shocks, all at the same velocities; hence with the present observations it is not possible to draw any statistically meaningful conclusion on the type of shock that is prevalent in outflows around low mass stars.

In conclusion, the H$_2$O/NH$_3$ as a function of velocity can be used to determine the most likely combination of 'pre-shock density and shock velocity', although it is not sufficient in itself to be able to constrain each individual parameter.

\end{enumerate}

\section{acknowledgements}
The Italian authors gratefully acknowledge the support from the Italian Space Agency (ASI) through the contract I/005/011/0, which also provided a fellowship for A.I G\'omez-Ruiz, who is now supported by Consejo Nacional de Ciencia y Tecnolog\'ia, through the program C\'atedras CONACYT para J\'ovenes Investigadores. I.J.-S. acknowledges the financial support received from the STFC through an Ernest Rutherford Fellowship (proposal number ST/L004801/1). HIFI has been designed and built by a consortium of institutes and university departments from across Europe, Canada and the United States under the leadership of SRON Netherlands Institute for Space Research, Groningen, The Netherlands and with major contributions from Germany, France and the US. Consortium members are: Canada: CSA, U. Waterloo; France: CESR, LAB, LERMA, IRAM; Germany: KOSMA, MPIfR, MPS; Ireland, NUI Maynooth; Italy: ASI, IFSI-INAF, Osservatorio Astrofisico di Arcetri-INAF; The Netherlands: SRON, TUD; Poland: CAMK, CBK; Spain: Observatorio Astron\'omico Nacional (IGN), Centro de Astrobiolog\'ia (CSIC-INTA). Sweden: Chalmers University of Technology - MC2, RSS \& GARD; Onsala Space Observatory; Swedish National Space Board, Stockholm University - Stockholm Observatory; Switzerland: ETH Zurich, FHNW; USA: Caltech, JPL, NHSC.

\bibliography{biblio}

\clearpage
\begin{appendix}
\section{Horizontal (H) and Vertical (V) polarization spectra}

The original vertical and horizontal polarization WBS spectra used to obtain the averaged spectra are shown in Fig. A1--A4. As mentioned in the observational section, the H and V beams were offset respect each other by $\sim$ 6.6$''$, which then produced that each polarization beam was covering a slightly different region. The effect of this offset can be seen in the separate H and V spectra. We notice that H and V are in agreement with the exception of IRAS4A-B, IRAS4A-R, and L1448-R4. An explanation in the case of the IRAS4A-B and IRAS4A-R is that, at each position, there is one beam (H or V) that covers more the central source than the other (see Fig. 1). In the case of L1448-R4, we see how the beam of V polarization covers more completely the bow of the R4 shock, while the H polarization beam misses this region.  

    \begin{figure*}
   \centering
      \includegraphics[angle=-90,width=15.0cm]{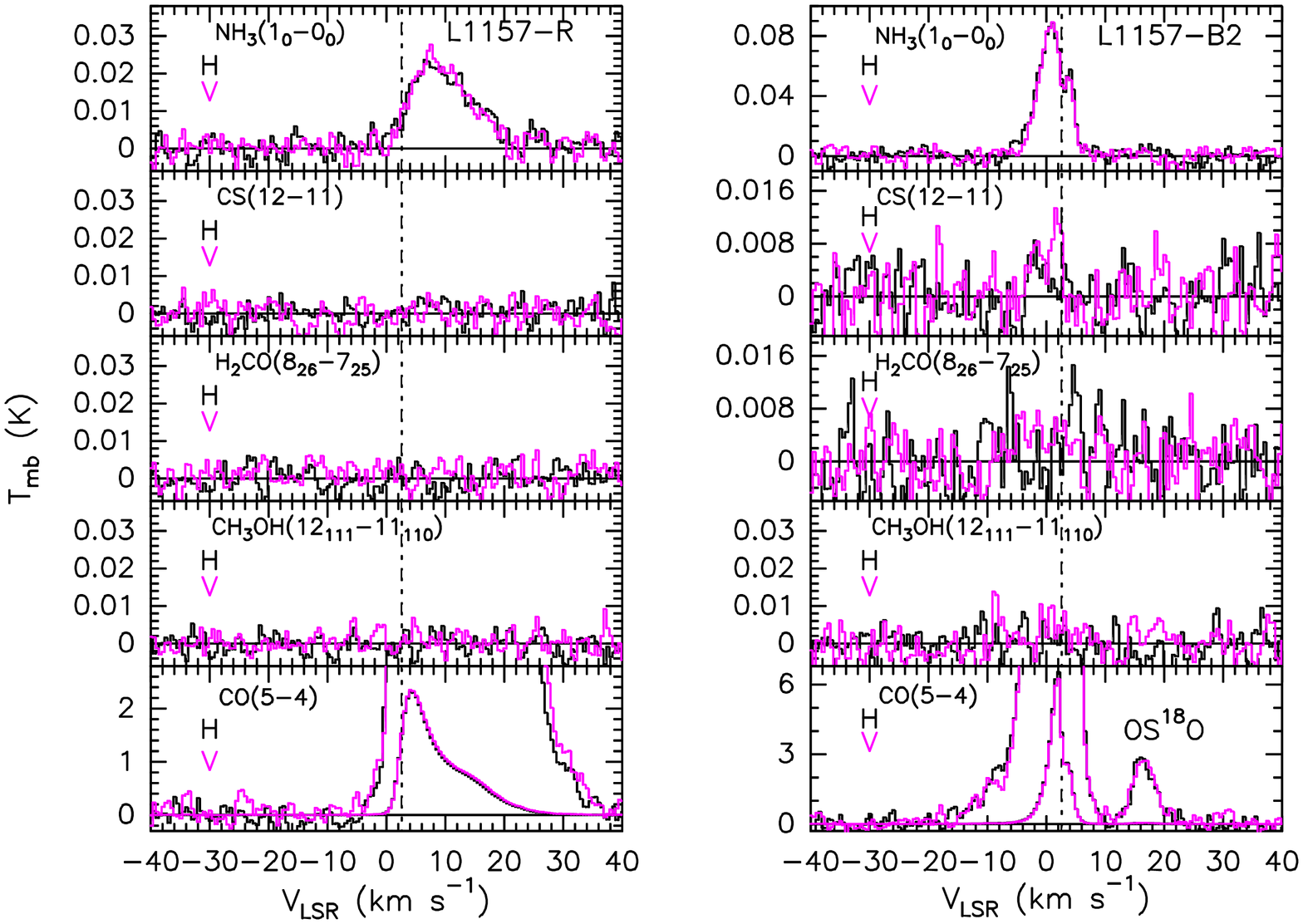}
      \caption{The NH$_3$, CS, H$_2$CO, CH$_3$OH, and CO transitions in L1557 positions (see Table 3). The WBS Horizontal (H) and Vertical (V) polarization spectra shown in black and magenta, respectively. Lower panels show also a zoom-in. Note that adjacent to CO (5--4) at B2, the OS$^{18}$O 7(5,3)--6(4,2) transition (E$_{\rm u}\sim$88 K, $\nu _0 =$ 576.24058 GHz) is also indicated. 
              }
         \label{l1157}
   \end{figure*}  

    \begin{figure*}
   \centering
      \includegraphics[angle=-90,width=15.0cm]{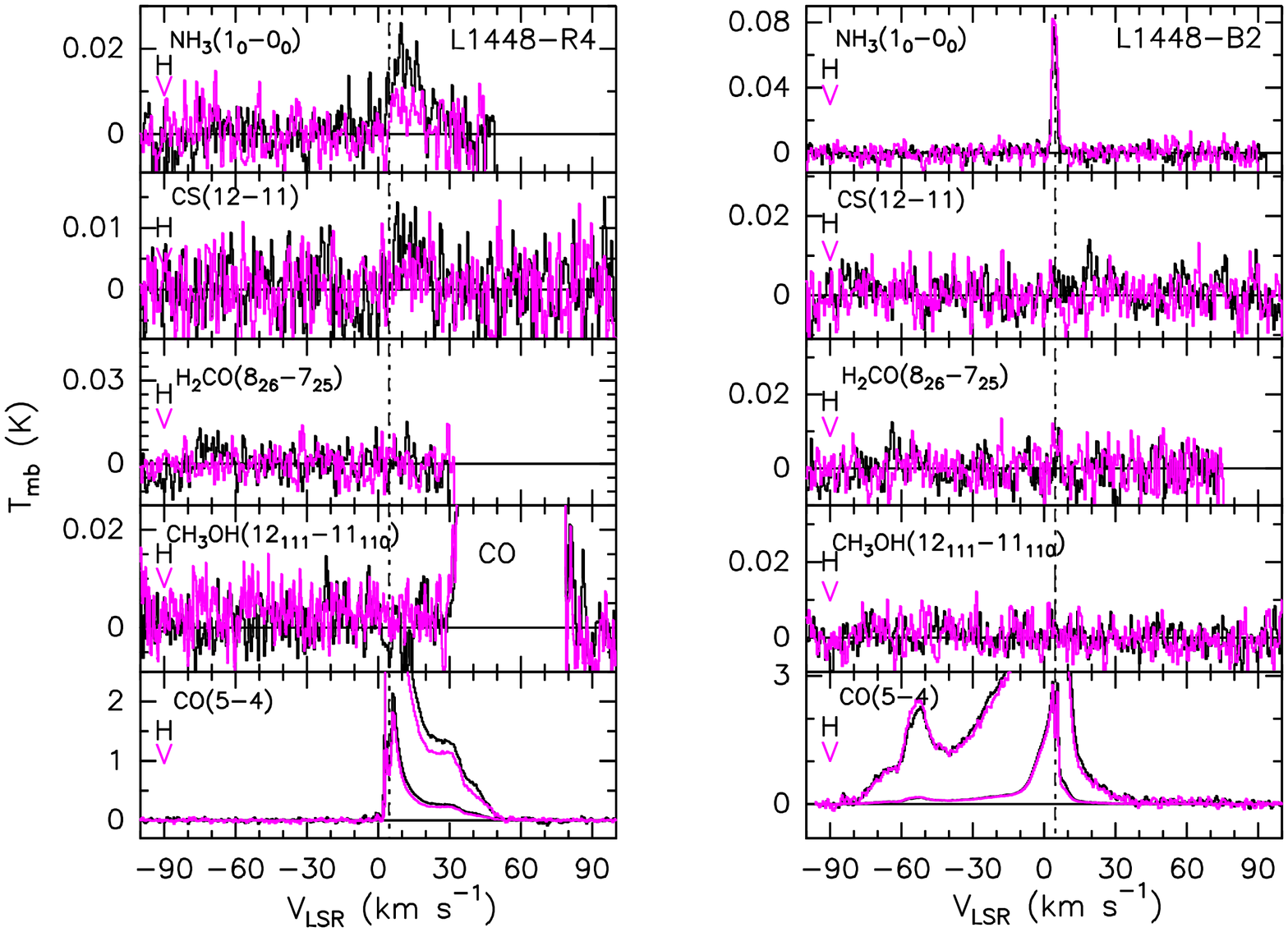}
      \caption{Same as Fig. \ref{l1157} but for L1448 positions.
              }
         \label{l1148}
   \end{figure*}

    \begin{figure*}
   \centering
      \includegraphics[angle=-90,width=15.0cm]{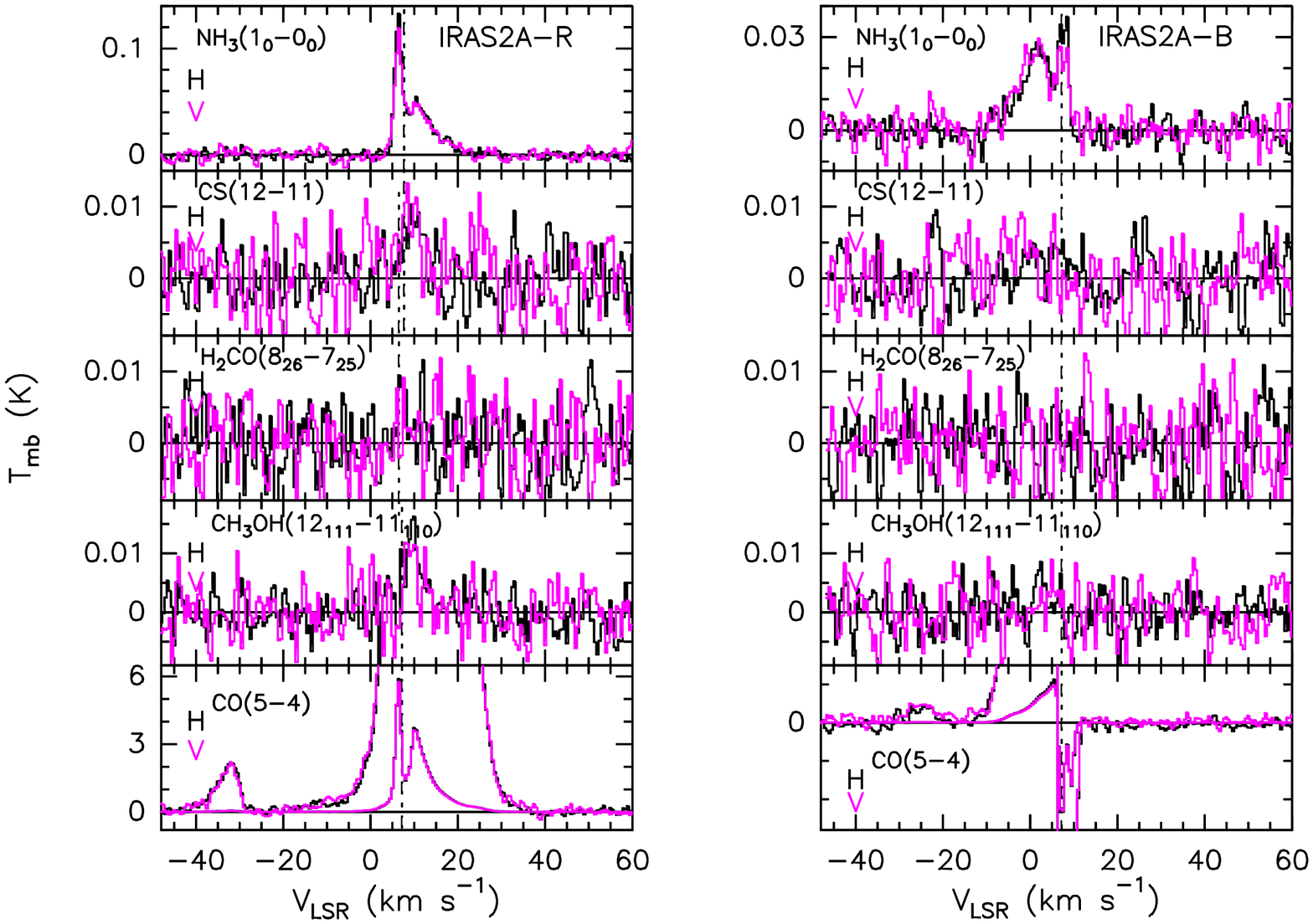}
      \caption{Same as Fig. \ref{l1157} but for IRAS2a positions. Note that contamination from the image band of the CH$_3$OH 6(1, 6)--5(0, 5) transition (E$_{\rm u}\sim$62 K, $\nu _0 =$ 584.4499 GHz) is seen at V$_{\rm lsr}\sim -$30 km s$^{-1}$.
              }
         \label{iras2}
   \end{figure*}   

    \begin{figure*}
   \centering
      \includegraphics[angle=-90,width=15.0cm]{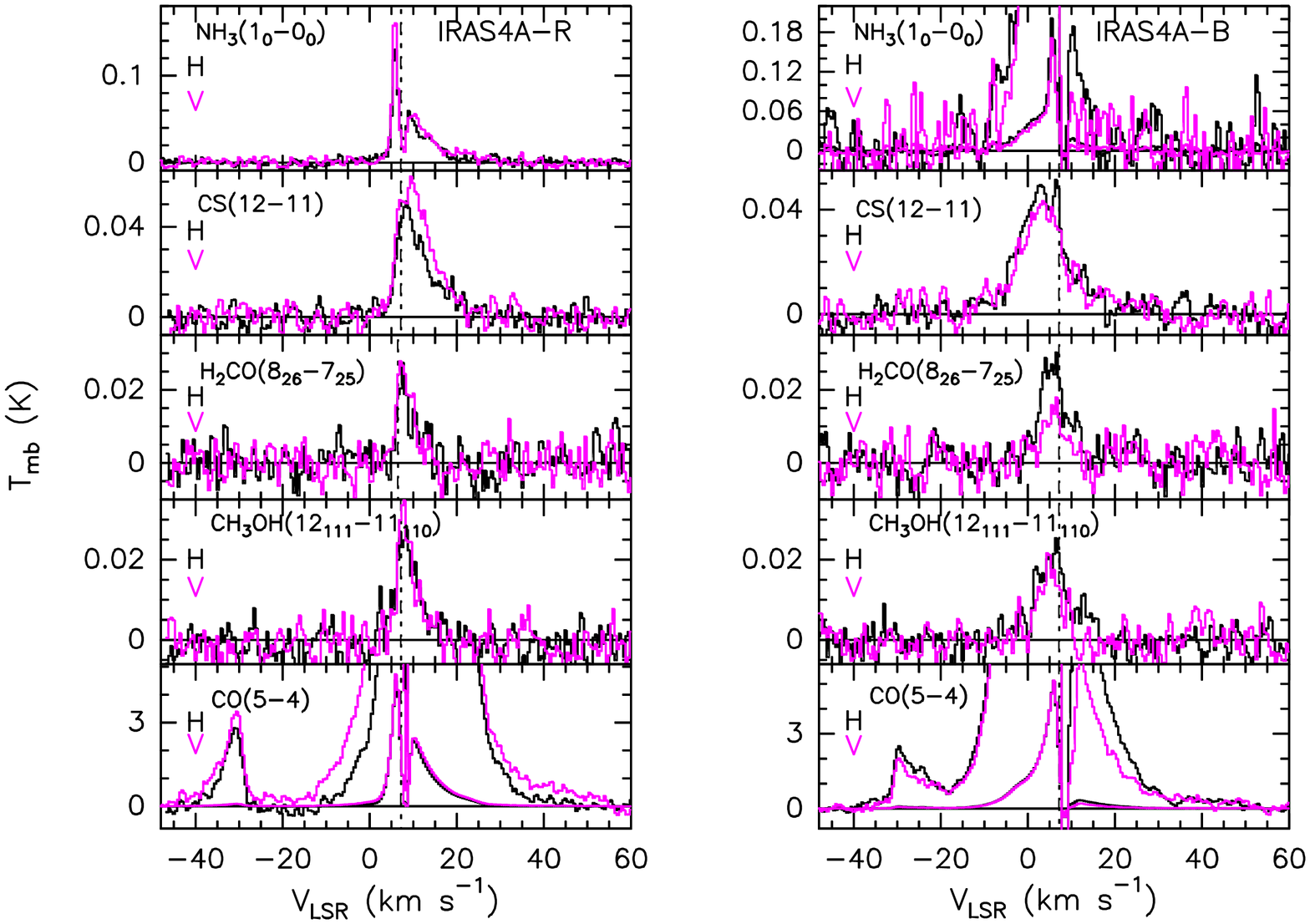}
      \caption{Same as Fig. \ref{l1157} but for IRAS4a positions. Note that contamination from the image band of the CH$_3$OH 6(1, 6)--5(0, 5) transition (E$_{\rm u}\sim$62 K, $\nu _0 =$ 584.4499 GHz) is seen at V$_{\rm lsr}\sim -$30 km s$^{-1}$.
              }
         \label{iras4}
   \end{figure*}

\section{Column densities and NH$_3$ opacity calculations}

We calculate the column densities of the observed species under the assumption of LTE and optically thin emission. The partition functions were calculated by using the standard assumptions \citep{Turner91} and with the molecular data taken from CDMS \citep{muller05}. The total integrated intensities reported in Table 4 are used. The following table shows the column densities for two temperatures, 20 K and 100 K.  

\begin{table*}
\centering
\caption{Column densities, from total integrated emission, assuming LTE and optically thin emission$^a$.}
\label{coldens}
\centering
\begin{tabular}{l cc c cc c cc c cc c cc}
\hline
Position & \multicolumn{2}{c}{o-NH$_3$ (1$_0$--0$_0$)} && \multicolumn{2}{c}{CO (5--4) $^b$} && \multicolumn{2}{c}{CH$_3$OH (12$_{1,11}$--11$_{1,10}$)A$^+$} && \multicolumn{2}{c}{p-H$_2$CO (8$_{2,6}$--7$_{2,5}$)} && \multicolumn{2}{c}{CS (12--11)} \\
 \cline{2-3} \cline{5-6} \cline{8-9} \cline{11-12} \cline{14-15} 
        & N(20) & N(100) && N(20) & N(100) && N(20) & N(100) && N(20) & N(100) && N(20) & N(100)\\
\hline
IRAS4A-B&	1.3 10$^{11}$ &	4.8 10$^{11}$ &&	6.5 10$^{16}$ &  1.2 10$^{16}$ && 7.4 10$^{15}$ &  3.1 10$^{13}$ && 9.1 10$^{13}$ &  9.8 10$^{11}$ && 5.2 10$^{14}$ &  1.7 10$^{12}$ \\
IRAS4A-R&	1.5 10$^{11}$ &	5.6 10$^{11}$ &&	6.2 10$^{16}$ &  1.1 10$^{16}$ && 8.1 10$^{15}$ &  3.4 10$^{13}$ && 8.1 10$^{13}$ &  8.7 10$^{11}$ && 4.4 10$^{14}$ &  1.4 10$^{12}$ \\
IRAS2A-B&	5.4 10$^{10}$ &	2.0 10$^{11}$ &&	1.2 10$^{16}$ &  2.2 10$^{15}$ && --     &  --    && --     &  --&&      --     &  --      \\ 
IRAS2A-R&	1.2 10$^{11}$ &	4.5 10$^{11}$ &&	7.3 10$^{16}$ &  1.3 10$^{16}$ && 1.9 10$^{15}$ &  7.8 10$^{12}$ && --     &  --&&      --     &  --    \\ 
L1448-B2&	4.1 10$^{10}$ &	1.5 10$^{11}$ &&	7.0 10$^{16}$ &  1.3 10$^{16}$ && --     &  --&&      --     &  --&&      --     &  --     \\
L1448-R4&	1.0 10$^{10}$ &	3.8 10$^{10}$ &&	4.0 10$^{16}$ &  7.3 10$^{15}$ && --     &  --&&      --     &  --&&      --     &  --     \\
L1157-B2&	9.0 10$^{10}$ &	3.7 10$^{11}$ &&	4.8 10$^{16}$ &  8.6 10$^{15}$ && --     &  --&&      --     &  --&&      --     &  --     \\
L1157-R	&	5.4 10$^{10}$ &	2.0 10$^{11}$ &&	4.8 10$^{16}$ &  8.7 10$^{15}$ && --     &  --&&      --     &  --&&      --     &  --     \\
\hline
\end{tabular}
\begin{center}
$^a$ The units are cm$^{-2}$, with N(20) and N(100) meaning the values obtained by assuming temperatures of 20 K and 100 K, respectively. $^b$ Due to effect of line absorption, this values should be considered as lower limits. \\
\end{center}
\end{table*}

\begin{figure*}
   \centering
      \includegraphics[angle=0,width=13.0cm]{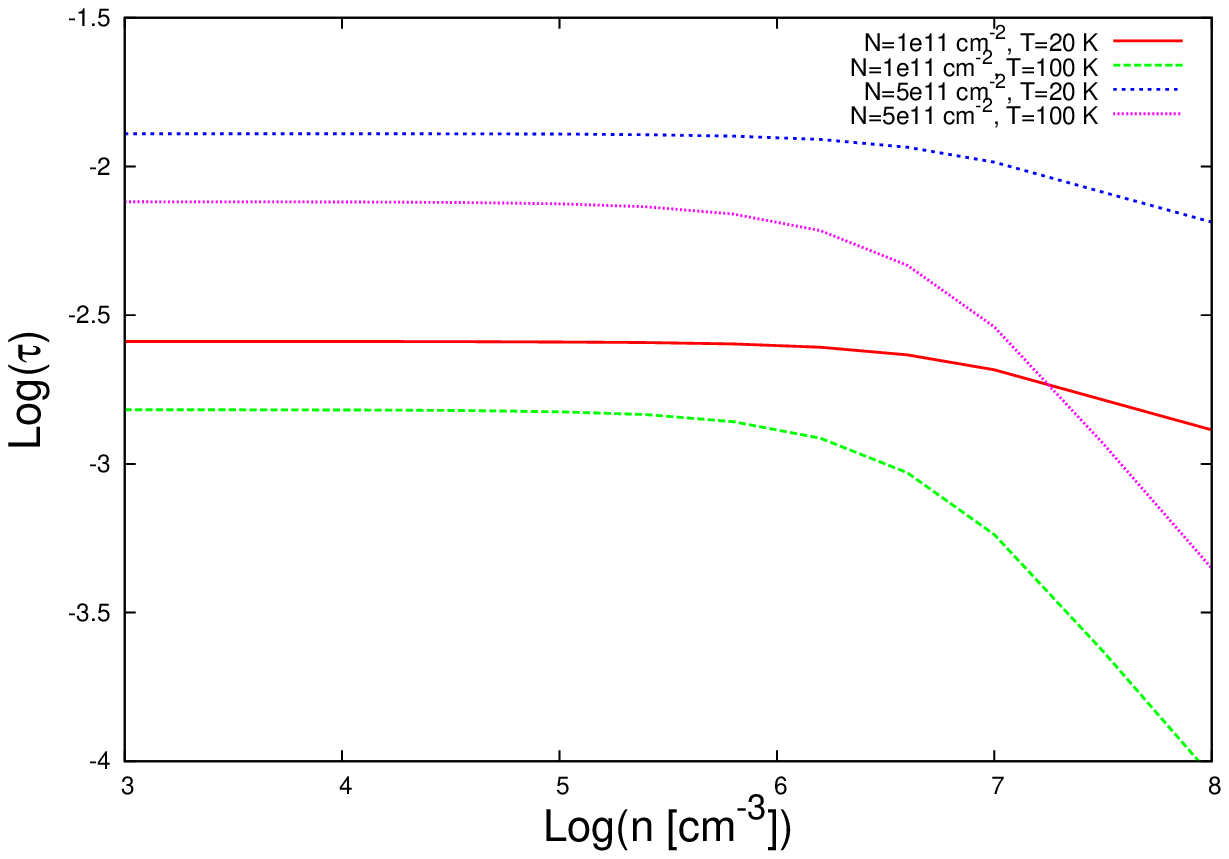}
      \caption{o-NH$_3$ (1$_0$--0$_0$) opacity as a function of H$_2$ particle density, from LVG calculations using RADEX. The different curves show different values of column density and kinetic temperature. Linewidth used in all cases is 10 km s$^{-1}$ (average observed linewidth). 
              }
         \label{tau-nh3}
   \end{figure*}

Using RADEX \citep{Tak07} we have computed the o-NH$_3$ (1$_0$--0$_0$) line opacity as a function of H$_2$ particle density. For the calculations we used an average linewidth of 10 km s$^{-1}$. The plots presented in Fig. \ref{tau-nh3} show the results for the column densities and temperatures values in Table \ref{coldens}. It is clear that in all cases $\tau$ $<<$ 1, also that $\tau$ is lower for higher H$_2$ particle density and kinetic temperature.

\end{appendix}

\end{document}